\documentclass[12pt]{article}

\usepackage[authoryear]{natbib}
\usepackage{amssymb}
\usepackage{amsthm}
\usepackage{amsmath, mathtools, bbm, dsfont}
\usepackage{subfigure} 
\usepackage{multirow}

\usepackage{xcolor}
\newcommand{\bphi}{\boldsymbol{\phi}}
\newcommand{\bq}{\boldsymbol{q}}
\usepackage{bbm}

\newcommand{\btheta}{\boldsymbol{\theta}}
\newcommand{\bd}{\boldsymbol{d}}
\newcommand{\bw}{\boldsymbol{w}}
\usepackage{setspace}
\newtheorem{lemma}{Lemma}

\usepackage{array}
\newcolumntype{L}[1]{>{\raggedright\let\newline\\\arraybackslash\hspace{0pt}}m{#1}}
\newcolumntype{C}[1]{>{\centering\let\newline\\\arraybackslash\hspace{0pt}}m{#1}}
\newcolumntype{R}[1]{>{\raggedleft\let\newline\\\arraybackslash\hspace{0pt}}m{#1}}
\usepackage{longtable}
\usepackage[hypertexnames=false,colorlinks,citecolor=blue,urlcolor=blue]{hyperref}

\usepackage{authblk}
\providecommand{\keywords}[1]{\textbf{\textit{Keywords---}} #1}

\begin{document}

\title{Statistical methods for linking geostatistical maps and transmission models: Application to lymphatic filariasis in East Africa}

\author[1,2]{Panayiota Touloupou}
 \author[3]{Renata Retkute}
\author[4]{T. D\'{e}irdre Hollingsworth}
\author[1,2]{Simon E. F. Spencer} 
\affil[1]{\small{Department of Statistics, University of Warwick, Coventry, UK}}
\affil[2]{Zeeman Institute, University of Warwick, Coventry, UK}
\affil[3]{Department of Plant Sciences,
University of Cambridge, Cambridge, UK}
\affil[4]{Big Data Institute, Li Ka Shing Centre for Health Information and Discovery, Nuffield Department of Medicine, University of Oxford, Oxford, UK}
\maketitle

\normalsize
\begin{abstract}

Infectious diseases remain one of the major causes of human mortality and suffering. Mathematical models have been established as an important tool for capturing the features that drive the spread of the disease, predicting the progression of an epidemic and hence guiding the development of strategies to control it. Another important area of epidemiological interest is the development of geostatistical methods for the analysis of data from spatially referenced prevalence surveys. Maps of prevalence are useful, not only for enabling a more precise disease risk stratification, but also for guiding the planning of more reliable spatial control programmes by identifying affected areas. Despite the methodological advances that have been made in each area independently, efforts to link transmission models and geostatistical maps have been limited. Motivated by this fact, we developed a Bayesian approach that combines fine-scale geostatistical maps of disease prevalence with transmission models to provide quantitative, spatially-explicit projections of the current and future impact of control programs against a disease. These estimates can then be used at a local level to identify the effectiveness of suggested intervention schemes and allow investigation of alternative strategies. The methodology has been applied to lymphatic filariasis in East Africa to provide estimates of the impact of different intervention strategies against the disease.
\end{abstract}


\section{Introduction}
\label{sec:Introduction}

Geostatistical modelling is increasingly used in epidemiology to combine surveys from multiple locations into a detailed model of local prevalence or incidence (\citealt{Hay:2009,Stensgaard:2011,Moraga:2015,OHanlon:2016}; \citealt{Giorgi2018}). Maps of disease distribution can be used, for example, to plan the development of national scale control strategies by informing policy makers where intervention efforts should be focused \citep{Tatem2010,Slater2013}. Several examples from the literature have shown that spatial heterogeneity is an important epidemiological factor in many diseases \citep[for example]{Sturrock:2010,Pullan:2012}, however predictions of future cases are frequently performed on aggregated data, risking the ecological fallacy \citep{Wakefield:2010}.

Over the past decades, mathematical models have also been established as an important tool for evaluating the effect of different control strategies by predicting the progression of the disease \citep{Ferguson2005,Tildesley2009,Stolk2018,hollingsworth2018}. However, when mathematical  modelling is used to evaluate potential intervention strategies, spatial heterogeneity is also frequently ignored \citep{Heesterbeek:2015}. Some notable exceptions are the papers by \citet{Gibson1997}, \citet{Keeling2001} and \citet{Deardon2010} who considered a spatial model, where the transmission probabilities depend on distances between individuals. In this paper we develop a novel method for taking the output from a geostatistical model and projecting the epidemic dynamics forward in time at the pixel level, under a range of potential intervention strategies, in a computationally efficient way. An important feature of our approach is the ability to capture several sources of uncertainty.

There are only a limited number of studies linking transmission models and geostatistical maps in a way that can dynamically inform policy at a local level. The African Program for Onchocerciasis Control was one of the first groups to develop and apply this approach (for example \cite{Plaisier:1991}, \cite{Alley:1994}). Kriging was used to extrapolate between survey points and then transmission models were used to project the likely impact of intervention programs. These mapped projections had been extremely useful in informing policy planning over many years and have recently been updated \citep{Tekle:2016}. The power of this type of approach to inform policy has been illustrated most notably by \cite{Bhatt:2015} in the analysis of the key drivers of successes in malaria interventions over the last 15 years. An important challenge, addressed by our approach, is to appropriately estimate and communicating the projections with their uncertainty. In particular, our method  addresses and quantifies a broad range of uncertainties, including uncertainty in the spatial variation in prevalence, transmission parameters, demographics, interventions and even model structure, and propagates them into the uncertainty in future predictions.

Our methodology has many parallels with exact versions of ABC \citep{Wilkinson2013,Beaumont2002}, in which simulations from the model are weighted (or accepted) according to their likelihood of producing the observed data. However, in our framework a likelihood is only available at the survey points, and so instead we weight the simulations by the posterior distribution from a geostatistical model that interpolates between surveys to give a prevalence distribution at each location. To achieve this weighting, we must change the measure of the simulated prevalences from the one induced by the prior on the transmission model parameters, to the posterior distribution from the geostatistical model using the Radon-Nikodym derivative \citep{Billingsley1995}. However, since this is not available in analytical form, we propose an empirical alternative similar to \citet{Goldie1999} and references contained within.

The paper is organised as follows. In Section \ref{sec:Methods} we describe the statistical methodology for combining geo-statistical mapping and transmission modelling, and illustrate its key features with a toy example in Section \ref{sec:toyexample}. The proposed method is applied in Section \ref{sec:ApplicationLF} to investigate the impact of intervention programs for lymphatic filariasis in seven countries in Africa. Finally, we conclude with a discussion on limitations of the current method and possible extensions for further research in Section \ref{sec:Discussion}.

\section{Methods}
\label{sec:Methods}

We develop a Bayesian methodology that captures uncertainty from multiple sources and can be readily applied to different transmission models and intervention strategies to give a distribution of projections across space. The starting point for our analysis is the output from a geostatistical model of disease prevalence, capturing the uncertainty in the spatial distribution of infection. A number of recent studies have adopted a predictive framework known as model-based geostatistics \citep{Diggle:2007} for the production of prevalence maps, often employing Bayesian inference for spatial prediction and robust characterisation of uncertainty surrounding those predictions. In particular we assume that the output consists of $M$ Monte Carlo samples from the posterior distribution of the geostatistical model. Although we assume that the spatial distribution represents the pre-control prevalence here, our methodology can easily be generalised beyond this example. In addition, we assume that other geographical information is available for each pixel (with associated uncertainty), such as population size and other demographic data that can be used as an input to the transmission model.

\begin{figure}[!]
\centering
	\includegraphics[scale=0.63]{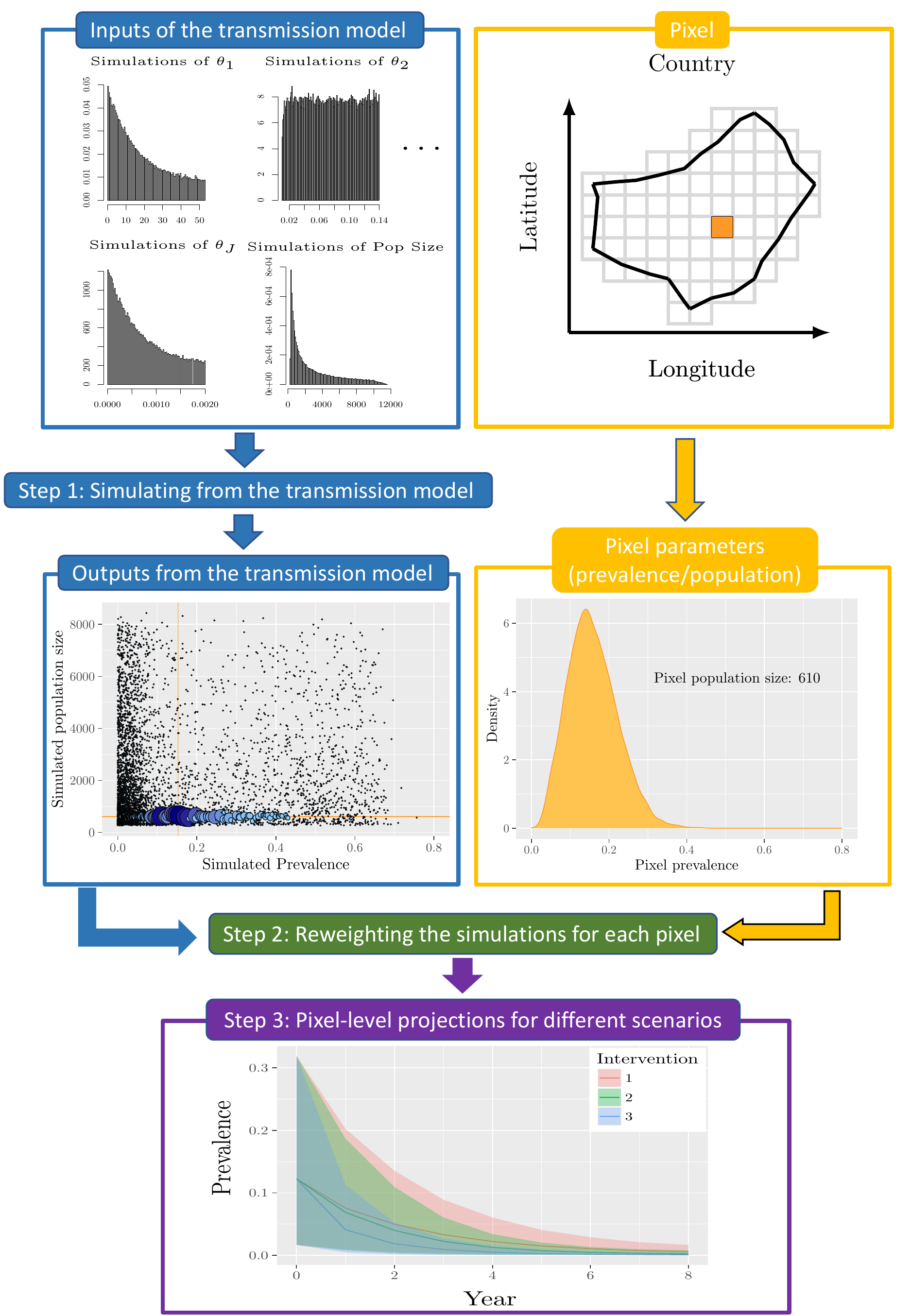}
\caption{Methodology for generating mapping results. Using pre-run model simulations (top), we reweight the simulations for each pixel based on the prevalence and population information (middle -- red lines represent the median of observed data).  Finally, the weights are used to evaluate the impact of different intervention strategies (bottom).}
\label{fig:Methods}
\end{figure}

Our methodology consists of 3 steps. First, we generate a large number of simulations from the transmission model, with sufficient variability to capture all of the endemic prevalences observed in the samples from the geostatistical model. Second, for each spatial location in the map we reweight the simulations according to how similar they are to the observed prevalence and other spatial information, such as population data. Finally, we simulate the transmission model further forward in time, possibly under some intervention strategy, and apply the weights to obtain the spatial distribution of the projections. A graphical representation of the method can be found in Figure \ref{fig:Methods}.

\subsection{Step 1: Simulating from the transmission model}
\label{sec:Methods:Step1}
For each pixel on the map, we assign an informative prior on the model parameters, $\pi_i(\btheta)$ say for pixel $i$, representing the uncertainty in our beliefs about the parameters of the transmission model at that location. Next, we define a single proposal distribution over the parameter space, $q(\btheta)$, capable of producing simulated prevalence levels spanning the values observed in the geostatistical mapping. We then draw $J$ parameter vectors ($\btheta_j$) from the proposal, and for each one we run the model forward in time until it reaches pre-control equilibrium. Denote the resulting prevalence levels by $p_j$, for $j=1,\dots J$. Finally, we calculate an initial $I\times J$ matrix of weights for the $I$ pixels and $J$ simulations according to the usual importance sampling formula, namely $w_{ij}^{(1)}=\pi_i(\btheta_j)/q(\btheta_j)$.

The proposal distribution $q(\btheta)$ might be uniform over the parameter space in low dimensions, or for higher dimensions it could be developed from pilot simulations, where parameter vectors are sampled uniformly from the support of the priors and for each parameter vector the equilibrium prevalence is simulated from the transmission model. The importance proposal can then be constructed on the parameter space to give more weight to frequently observed prevalence values, and zero weight to implausible prevalence values (e.g.~prevalences larger than the maximum observed in the geostatistical model). The efficiency of the proposal can be improved iteratively using adaptive importance sampling techniques \citep{Cornuet:2012}.

\subsection{Step 2: Reweighting the simulations to match pixel prevalence distributions}\label{step2}

For each pixel the same simulations are reweighted to match the prevalence distribution of that pixel. This prevents unnecessary replication of simulations for pixels that are broadly similar and means that the number of simulations  need not increase as the number of pixels increases. More specifically, for pixel $i = 1,\ldots, I$ and simulation $j = 1,\ldots, J$  the new weight is given by:
\begin{equation}  
  w_{ij}^{(2)} \propto \frac{f(p_j|\bd_i)}{g(p_j|\bw_i^{(1)})}w_{ij}^{(1)}
  \label{eq:Methods1}
\end{equation}
where $\bd_i=(d_{i1},\dots,d_{iM})$ is the $M$ dimensional vector of posterior samples of prevalences in pixel $i$ and $\bw_i^{(1)}=(w_{i1}^{(1)},\dots,w_{iJ}^{(1)})$. The function $f$ represents the probability of having prevalence $p_j$ under the geostatistical model and $g$ represents the probability of simulating prevalence $p_j$ from the model with parameter vector drawn from the prior. The ratio $f/g$ therefore represents the usual change of measure formula (Radon-Nikodym derivative). However, since neither of these probability densities are likely to be available in closed form, we use an empirical approximation given by the amount of probability density within $\delta/2$ of $p_j$: 
\begin{align*}
f(p_j|\bd_i)&= \displaystyle \frac{1}{\delta M}\sum\limits_{m=1}^{M} \mathbbm{1}_{ \left \{ p_j  -  {\delta/2} \, \leq \, d_{im} \, \leq \, p_j  +  { \delta/2} \right\} },\\
g(p_j|\bw_i^{(1)})&= \displaystyle \frac{\sum_{k=1}^{J} w_{ik}^{(1)}\mathbbm{1}_{\{ p_j  -  {\delta/2}\, \leq \, { p_k} \, \leq \, p_j  +  { \delta/2} \} }}{\delta \sum_{k=1}^{J} w_{ik}^{(1)}}.
\end{align*}

Note that as long as $q(\btheta)>0$ implies $\pi_i(\btheta)>0$, then $w_{ij}^{(1)}>0$ for all $j$ and hence $g(p_j|\bw_i^{(1)})>0$ for all $j$.  The bin width $\delta$ controls the trade-off between effective sample size and the fidelity of the distribution of the simulated prevalences to the geostatistical posterior distribution, and should be set as small as possible, whilst producing a reasonable effective sample size, defined as $ \left. {\left(\sum _{j=1}^{J}w^{(2)}_{ij}\right)^{2}} \middle/ {\sum _{j=1}^{J} \left(w^{(2)}_{ij} \right)^{2}} \right.$. Finally, the weights from Equation \eqref{eq:Methods1} are normalised to give the posterior probabilities (according to the geostatistical model) that simulation $j$ is appropriate for pixel $i$. 

\subsection{Step 3: Running the simulations forward}

The simulations are run forward in time under a given intervention strategy. For each pixel the projected outcomes are weighted according to the normalised weights $\bw_i^{(2)} = (w_{i1}^{(2)},\dots,w_{iJ}^{(2)})$ produced in Step 2. Step 3 is repeated for each intervention strategy under consideration.

\subsection{Lemma on the change of measure}

In this section we introduce a lemma that generates the reweighting formula in Equation \eqref{eq:Methods1}. The lemma is proved in Section \ref{proof} of the Appendix. 

\begin{lemma}\label{lemma}
Let $p:\mathbb{R}^d \rightarrow [0,1]$ denote a deterministic model that produces a prevalence $p(\btheta)$ from a vector of parameters $\btheta$. Let $\pi(\btheta)$ be a prior distribution over the parameters that induces a prior distribution over prevalences, which we denote by $g(p)$. Suppose that there exists a differentiable and invertible function $\bphi: \mathbb{R}^d\rightarrow\mathbb{R}^d$ that admits the prevalence as its first argument, ie. $\bphi(\btheta)=(p(\btheta),\bq(\btheta))$ for some $\bq(\btheta)$. Finally suppose that we wish to change the probability measure over prevalences from $g$ to another measure $f$ that is absolutely continuous with respect to $g$. Then the resulting measure over the parameter space is given by $h(\btheta)=\frac{f(p(\btheta))}{g(p(\btheta))}\pi(\btheta).$
\end{lemma}

\textbf{Notes:}
\begin{enumerate}
    \item The same approach can be applied for stochastic transmission models as long as the model is defined on a separate probability space $(\Omega,\mathcal{F},P)$ to the prior. For stochastic models we fix $\omega\in\Omega$ and consider the transmission model as a deterministic map $\bphi(\btheta,\omega)$, applying the Lemma and then integrating over $\Omega$.
    \item The condition that $f$ must be absolutely continuous with respect to $g$ means that whenever $g(p)=0$ then we must also have $f(p)=0$. In other words, when the prior probability of a prevalence is zero then the map measure of prevalence must also be zero. This has important implications for the implementation of our method, discussed further in Section \ref{sec:SM_note2}. 
\end{enumerate}

\subsection{Alternative empirical Radon-Nikodym derivatives}
\label{sec:AlternativeERND}

In Step 2 of our algorithm (described in Section \ref{step2}) we proposed an empirical estimate of the Radon-Nikodym derivative $f/g$ based on using the prevalences within $\delta/2$ of $p_j$. Clearly there are many possible alternative estimates that could be used and there are two in particular that are worthy of further discussion. The first is based on histograms and the second is based on minimising a discrepancy measure.

\subsubsection{Histogram-based empirical Radon-Nikodym derivative}
\label{sec:histogram}

If we consider a fixed partition of the prevalence space into bins (as if we were constructing a histogram) then it is straightforward to calculate the Radon-Nikodym derivative $f/g$ for each bin as the proportion of posterior samples in the bin divided by the proportion of the weight belonging to simulations that fall in the bin. More precisely, given a finite set of disjoint intervals with union $[0,1]$ then if prevalence $p_j$ falls in interval $\mathcal{I}(p_j)$ we have that
\begin{align*}
f(p_j|\bd_i)&= \displaystyle \frac{1}{M|\mathcal{I}(p_j)|} \sum\limits_{m=1}^{M} \mathbbm{1}_{\{d_{im}\in\ \mathcal{I}(p_j)\}},\\
g(p_j|\bw_i^{(1)})&= \displaystyle  \dfrac{\sum_{k=1}^{J} w_{ik}^{(1)}\mathbbm{1}_{\{p_k \in\  \mathcal{I}(p_j)\}}}{|\mathcal{I}(p_j)|\sum_{k=1}^{J} w_{ik}^{(1)}},
\end{align*}
where $|\mathcal{I}(p_j)|$ is the length of the interval containing $p_j$.

The main advantage of this estimate is computational -- since all of the simulations in the same interval have the same ratio (for a given pixel) then instead of having to calculate $J$ ratios we need only calculate one per interval. A secondary advantage is that the weighted histogram of the simulation prevalences will be identical to the histogram of the posterior prevalence distribution. However, the relative weightings within each bin are unchanged and so a different choice of bins will reveal that the two distributions are different.

\subsubsection{Discrepancy-based empirical Radon-Nikodym derivative}
\label{sec:discrepancy}

A second alternative empirical Radon-Nikodym derivative can be defined to minimise the difference between the empirical cumulative distribution functions (cdfs) of the posterior prevalences and the weighted simulated prevalences. Let $F(x|\bd_i)=\frac{1}{M}\sum_{m=1}^M\mathbbm{1}_{\{d_{im}\leq x\}}$ be the empirical cdf of the map prevalence distribution for pixel $i$ and $H(x|\bw^{(2)}_i)=\sum_{j=1}^Jw^{(2)}_{ij}\mathbbm{1}_{\{p_j\leq x\}}$ be the empirical cdf of the final weighted distribution of simulated prevalences, then we can choose $\bw^{(2)}_i$ to minimise some distance $||F(\cdot|\bd_i)-H(\cdot|\bw^{(2)}_i)||$. For example, we may wish to minimise $\int_0^1 |F(x|\bd_i)-H(x|\bw^{(2)}_i)|\:\mathrm{d}x$ or $\allowbreak  \int_0^1 \left(F(x|\bd_i)-H(x|\bw^{(2)}_i)\right)^2\:\mathrm{d}x$. In this paper we have focussed on the latter of these, for details of the calculation we refer the reader to the Appendix \ref{sec:SM_Minimum}.

\section{Simulation studies: A toy example}
\label{sec:toyexample}

In this section we provide a toy example to assess the performance of the proposed method under different settings. Particular focus was given on how the method was affected by the value of $\delta$, by the choice of the proposal distribution of the parameters and the empirical estimate of the Radon-Nikodym derivative. A full description of the analysis can be found in Appendix  \ref{sec:Toyexample} and here we summarize the key results.

Suppose that the prior distribution is $\pi(\theta_1,\theta_2)=2$ if $0<\theta_2<\theta_1<1$ and zero otherwise. Plots illustrating this prior are given in Appendix Figure \ref{fig:toyPrior}. For simplicity, assume that the transmission model has equilibrium prevalence given by $p(\theta_1,\theta_2)=\theta_1$ so that the induced prior over prevalences is the marginal for $\theta_1$, ie.~$g(p)=2p$ for $0<p<1$, which is a Beta(2,1) distribution. Further, suppose that we are given $M=2000$ samples from a pixel with prevalence measure $f(p)=2(1-p)$ for $0<p<1$, representing a Beta(1,2) distribution. This challenging example allows us to assess how the methodology performs when there are few simulations with low weights in the area of high posterior probability close to $p=0$.

\begin{figure}[!]
\centering
\subfigure[The proposal yields a simulated prevalence distribution equal to its marginal prior, Beta(2,1).\label{fig:BestDeltaToyPrior}]{
\includegraphics[scale=1]{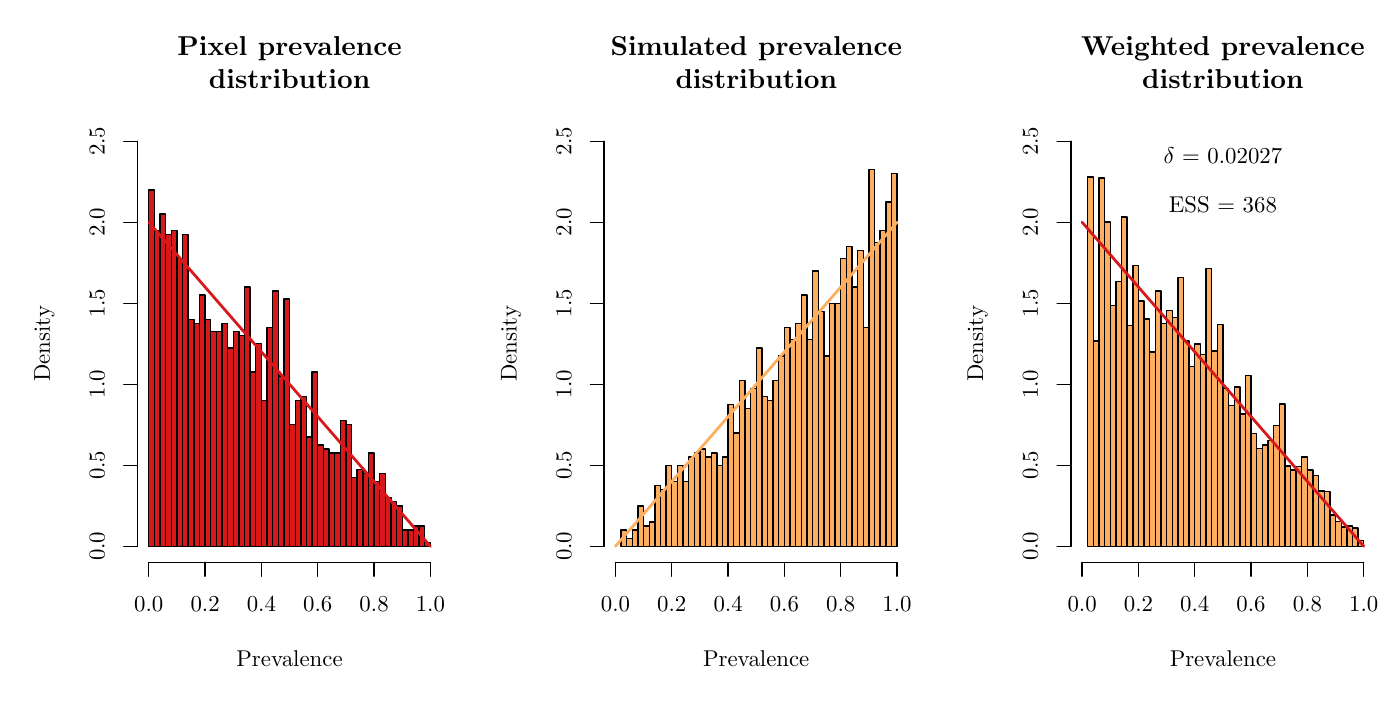}
}
\subfigure[The proposal yields a uniform simulated prevalence distribution. \label{fig:BestDeltaToyUniform}]{
\includegraphics[scale=1]{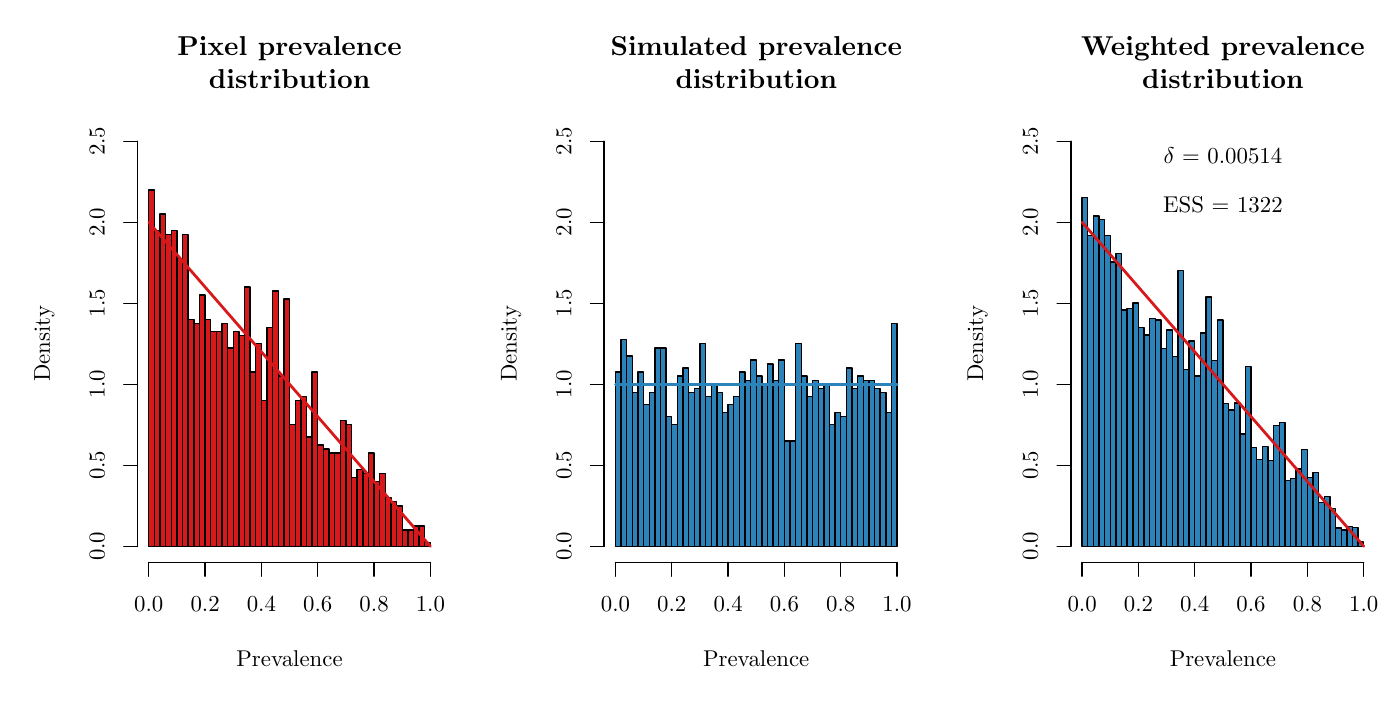} 
}
\caption{The estimated weighted prevalence distribution for the suggested value of $\delta$ (right panel) is compared to the true pixel prevalence distribution (left panel), under different proposal distributions for the prevalence (middle panel): (a) Beta(2,1) and (b) $U(0,1)$. The target densities are also shown on each panel. \label{fig:BestDelta}}
\end{figure}

Simulations were conducted to investigate the accuracy and efficiency of the proposed method under different settings, where the observed pixel and simulated prevalence data are obtained from the toy model. Figure \ref{fig:BestDelta} shows how $J=2000$ simulations from a proposal (centre histogram) can be reweighted (right histogram) to resemble the pixel prevalence distribution (left histogram). In Figure \ref{fig:BestDelta}(a) the proposal is from the prior, whilst in Figure \ref{fig:BestDelta}(b) the proposal is $U(0,1)$. The improvement due to the proposal distribution having good support in all areas of the posterior distribution was demonstrated by the substantial increase in effective sample size (ESS; from 368 to 1322), despite a much smaller value of $\delta$.

\begin{figure}[h!]
\centering
\includegraphics[scale=1]{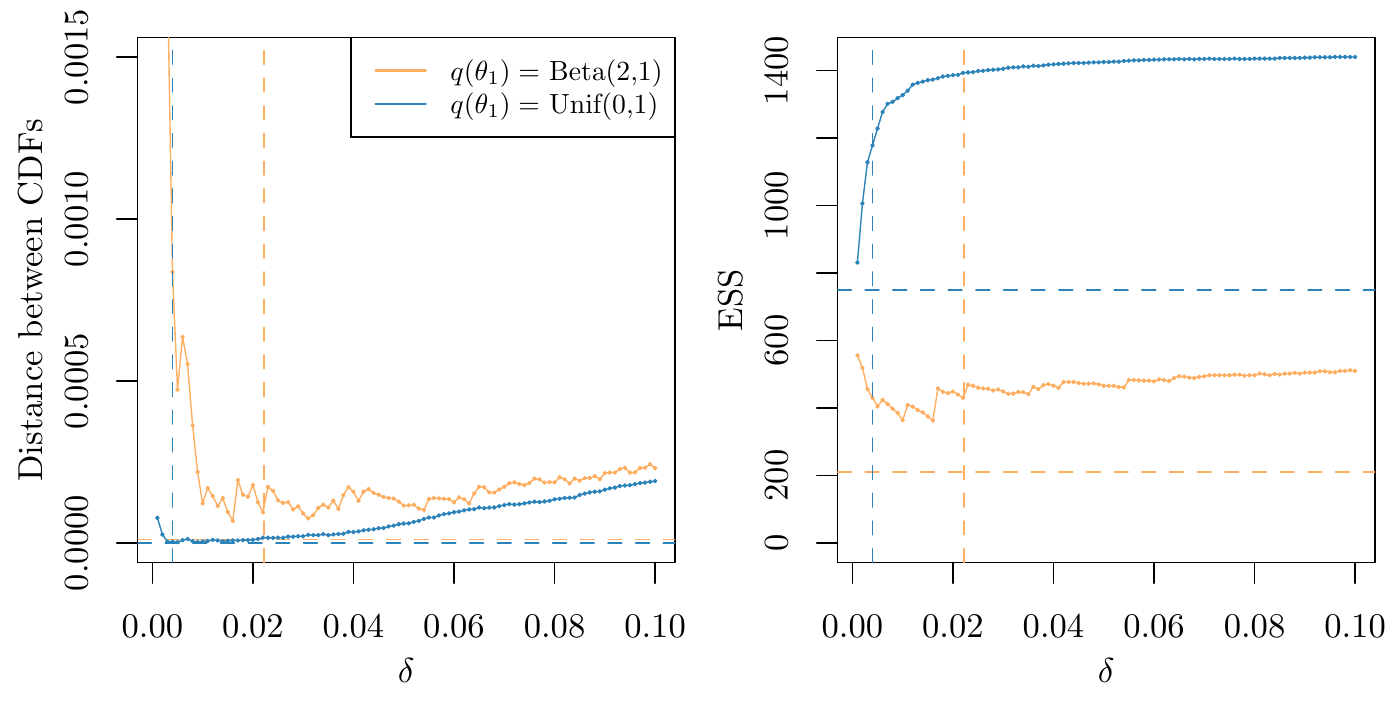}
\caption{Distance between the two cumulative distribution functions (cdfs) (left panel) and effective sample size (ESS) (right panel) obtained under different values of $\delta$ and choice of proposal distribution for parameter $\theta_1$, for one randomly selected simulated dataset. Orange solid line represents a prevalence proposal distribution equal to the marginal prior, i.e. Beta(2,1), whereas the blue line corresponds to a $U(0,1)$ proposal. In both cases, the pixel prevalences were drawn from a Beta(1,2) distribution. Dashed vertical lines represent the suggested value of $\delta$ for each scenario considered. Dashed horizontal lines correspond to the minimum possible distance (left panel) and its associated ESS (right panel).}
\label{fig:Distance_ESS_Toy_1Simul}
\end{figure}

Figure \ref{fig:Distance_ESS_Toy_1Simul} illustrates how the performance changes as $\delta$ is increased. The left figure shows the distance (given by integrated squared difference) between the empirical cumulative distribution functions of the weighted simulations and the samples from the pixel posterior; and the right figure shows the effective sample sizes. The corresponding results from the discrepancy based empirical Radon-Nikodym derivative (see Section \ref{sec:discrepancy}) are shown as horizontal dashed lines and provide the minimum distance possible between the cdfs. The results show that smaller values of $\delta$ reproduced the empirical cdf more accurately, unless $\delta$ was so small that very few simulations were included in each estimate of the density $g$ (the denominator in Equation \eqref{eq:Methods1}). After some experimentation (see Appendix   \ref{toydelta}) we chose to set $\delta$ to be the smallest value for which at least three simulations were included in each estimate of $g$. These values are illustrated by vertical lines on Figure \ref{fig:Distance_ESS_Toy_1Simul}, and can be seen to come close to achieving the minimum possible squared distance between the empirical cdfs, but with larger ESS. 

In our simulations we have evaluated the performance of the method for the distance-based empirical Radon-Nikodym derivative, which is based on using the prevalences within a certain distance given by $\delta/2$ of $p_j$. We also investigated alternative derivatives, discussed in Section \ref{sec:AlternativeERND}, and the results are summarised in Table \ref{tab:ERND} of the Appendix. Overall, we observed that the discrepancy-based derivative provides the best possible distance between the two cdfs, but at a cost of a lower ESS in all the scenarios considered. When the proposal was uniform and had simulations in all areas of the posterior distribution then the histogram-based derivative performed better than the distance-based derivative both in terms of accuracy and ESS. However, the situation was reversed when there were areas in the proposal with few simulations. In that scenario, the distance-based derivative was found to have lower integrated squared distance and higher ESS compared to the histogram-based derivative. Therefore, we used the distance-based empirical Radon-Nikodym derivative in our applications, since it was more robust to weaknesses in the proposal.

\section{Application to lymphatic filariasis data}
\label{sec:ApplicationLF}

In this section, we apply the proposed approach for the analysis of real data for lymphatic filariasis (LF) in East Africa. LF is caused by a mosquito-borne macro-parasite, which was historically endemic in many tropical countries, with over a billion people at risk of infection, and millions affected by the disease suffer from disability, stigma and associated social and economic consequences \citep{Ramaiah:2014}. LF is one of the neglected tropical diseases (NTDs) targeted for elimination as a public health problem by 2020 \citep{WHO:2012}, with new guidelines currently being developed for 2030.
Global efforts to eliminate LF as a public health problem, through the use of mass drug administration (MDA) of treatments with an excellent safety record, have reduced prevalence to low levels in many settings \citep{Ramaiah:2014}. While many countries have successfully scaled-up their programs, there remain a number of questions on how best to scale up treatment to assist priority countries in optimising interventions to accelerate elimination. Therefore, there is an urgent need to provide detailed estimates of the impact of current and future control programs for donor and policy planning.

For LF, the intervention strategy for most of Africa is to have yearly MDA at 65\% coverage for 5 years, followed by an assessment of transmission and, if necessary, further rounds of treatment. In areas where MDA has not yet started, alternative strategies may be required to meet the WHO target, i.e. the prevalence being less than 1\% \citep{WHO:2012} as soon as possible. Enhanced strategies include MDA at high coverage or twice-yearly treatment \citep{Stolk:2018}. By bringing together statistical mapping and transmission modelling, we aim to provide high-resolution quantification of the likely impact of control programs and predictions on both future impact and demand for interventions, allowing policy makers to more effectively target available resources. 

\subsection{The mathematical model of LF transmission dynamics}
\label{sec:Model}

In this section we describe the mathematical model of lymphatic filariasis transmission, TRANSFIL \citep{Irvine:2015}, that is used throughout the paper. TRANSFIL is an individual-based model of LF infection in human populations, with each host having their own adult worm and microfilariae (mf) burden, as well as mosquito bite risk and treatment history. A full description of the model is provided in \cite{Irvine:2015} and in Section \ref{sec:TRANSFIL} of the Appendix, so here we provide only a brief overview and the updated aspects of it.

Each human is assumed to have their own burden of male and female worms denoted by $W_i^m$ and $W_i^f$, respectively. The times at which human $i$ acquires female and male adult worms are given by two inhomogenous Poisson processes, both with rate:
 \begin{equation*}
 \tfrac{1}{2} \lambda b_i (V/H) \psi_1 \psi_2 s_2 h(a),
 \end{equation*}
where $\lambda$ is the number of bites per mosquito, $V/H$ is the ratio of vectors to hosts, $\psi_1$ is the probability that a third-stage larvae (L3, the infectious stage) leaves the host during a bite, $\psi_2$ is the probability that the L3 enters the host, $s_2$ is the proportion of L3 that develop into adult worms within the host and $h(a)$ is the biting rate for a human with age $a$. Both male and female worms are introduced to a human according to a bite risk $b_i$ drawn from a gamma distribution with mean 1 and shape parameter $k$.
Thus, the degree of parasite aggregation amongst humans can be quantified by this shape parameter. Finally, we assume that each worm has a constant death rate $\mu$.
 
Microfilariae concentration in the peripheral blood, denoted by $M_i$, is also modelled for each individual according to the following equation:
 \begin{equation*}
   \frac{\text{d}M_i}{\text{d}t} = \alpha W_i^f \mathbbm{1}_{\{W_i^m>0\}} - \gamma M_i,
 \end{equation*}
with $\alpha$ being the production rate of mf per worm, $\gamma$ the constant death rate of mf and the indicator function $\mathbbm{1}_{\{W_i^m>0\}}$ is one if there are male worms and zero if not. Larvae development occurs when mf entered the mosquito during a blood meal from an infected host. Different functional forms have been found to describe the relationship between the number of mf ingested and the number that develop within the mosquito. For \textit{Anopheles}, which is the genus of the most dominant vector species in East Africa, this relationship  is expressed as:
 \begin{equation*}
 L(m)  = \kappa_{s2}\left( 1 - e^ {-r_2m/\kappa_{s2}} \right)^2, 
 \end{equation*}
where $m$ is the concentration of mf per $20\mu$L taken during a blood meal and $r, \kappa_s$ denote the saturation values related to the uptake function as detailed in \cite{Gambhir:2008}. The equilibrium value for L3 in a mosquito is given by:
\begin{align*}
L^* = \frac{\lambda g \tilde{L} }{\sigma + \lambda \psi_1},
\end{align*}
 where $\lambda$ is the number of bites per mosquito, $g$ is the proportion of mosquitoes which pick up infection when biting an infected host, $\sigma$ is the death rate of mosquitoes and $\tilde{L}$ is the average number of larvae per mosquito.

Each human begins life with zero infection and a bite-rate of exposure $b_i$. The human death rate is denoted by $\tau$ and is assumed to be constant throughout an individual’s lifetime with a cut-off at age 100. When an individual dies another one is born in order to keep the population size constant.

During an intervention campaign, the impact of MDA is simulated for an individual by reducing their mf concentration and their male and female worm burden according to the estimated drug efficacies from the literature \citep{Ismail:1998,Michael:2004}. In addition, there is a period after MDA during which the production of mf for that individual is diminished. Furthermore, the individuals' compliance after multiple rounds of treatment is modelled based on the paper by \cite{Griffin:2010}, where the authors model the probability of an individual making the same decision as in the previous round of treatment.

Finally, we extended the model to include a very low rate of importation of infection from outside the population being modelled, otherwise the equilibrium distribution (steady state), that is used as
the starting point of the simulations, is just the degenerate distribution where no-one is infected. The interventions reduce the prevalence over time, and so we reduce the importation rate after intervention in proportion to the reduction in prevalence seen in pilot simulations. Lists of the model parameters are provided in Tables \ref{tab:Parameters} and \ref{tab:ParametersVary} of the Appendix.

\subsection{Implementation details}
\label{subsec:Implementation}

The starting point for our analysis was the spatial map (pixel scale $5\times5$ km) providing the predicted distribution of the LF prevalence based on mf data, generated through a Bayesian geostatistical modelling approach described by \cite{Moraga:2015}. The top panel of Figure \ref{fig:ObservedVsEstimated} shows the median of the posterior distribution of the prevalence obtained at each pixel, along with estimates of lower (2.5\%) and upper (97.5\%) quantiles. In particular, we analysed the following seven African countries: Ethiopia, Sudan, South Sudan, Eritrea, Kenya, Tanzania and Uganda. We linked each pixel to the corresponding population estimates obtained from the Gridded Population of the World \citep{Worldpop}, which provides the estimated number of people in each pixel. 
We avoided handling pixels with either very small or very large populations as the transmission model was not thought to be appropriate in these environments  \citep{Irvine:2015,Smith:2017}. More specifically, for small populations we pooled pixels with less than 300 people together, ensuring that the merged pixels belong to the same country and that the groups contain as few pixels as possible. We excluded pixels with population estimates over 10\,000 from the analysis; resulting in 1.7\% of the pixels being excluded.

The stochastic model of LF transmission TRANSFIL was used to investigate and compare the impact of different control strategies. In order to simulate the entire range of observed baseline mf prevalence levels, with values up to 95\%, we assumed that four parameters of the mathematical model were spatially varying: the population size, the vector to host ratio, the aggregation parameter of individual exposure to mosquitoes and the importation rate, using prior distributions informed from data, pilot simulations and previous analyses. We assumed that the parameter prior was the same in each spatial location (discussed in more details in Section \ref{sec:TRANSFIL:Implementation} of the Appendix) except for the population size, which was assumed to be a log normal distribution, ie  $\log(n) \sim \mathcal{N}\left(\log(N_i), \sigma^2 \right)$, where $N_i$ is the reported population of pixel $i$ \citep[adjusted population from][]{Worldpop} and $ \sigma$ is the sample standard deviation of the log population estimates available in WorldPop.

The proposal density of the population sizes, $q(n)$, was designed so that each simulation contributed an equal amount to the effective sample size of a set of pixels with populations $\{260,261,\dots,10\,000\}$. This was achieved by calculating the  effective sample size of an initial proposal, namely, $q_0(n)\propto 1$. The remaining populations sizes $(10\,001$--$11\,550)$ were taken to decrease linearly from $q(10\,000)$ to zero. Since the uncertainty in the log-normal prior is much greater for large populations, fewer simulations are needed in these regions. The final proposal was obtained from 10 iterations of $q_i(n)\propto q_{i-1}(n)/\text{ESS}_{i-1}(n)$, where $\text{ESS}_i(n)$ is the effective sample size of a simulation with $n$ individuals from the proposal $q_i$ (see Figure \ref{fig:Distributions}(a) of the Appendix).

A significant merit of our approach is that it can be easily applied in parallel which can be utilised to speed up implementation, especially in applications involving a large number of pixels. This is because we are treating each pixel independently and therefore the computation of the weights can be undertaken in parallel. In our application, the computation time of this step was approximately 8 hours using a 112 core computer cluster.

\subsection{Results}
\label{subsec:Results}

\begin{figure}[!]
\centering
	\includegraphics[scale=1]{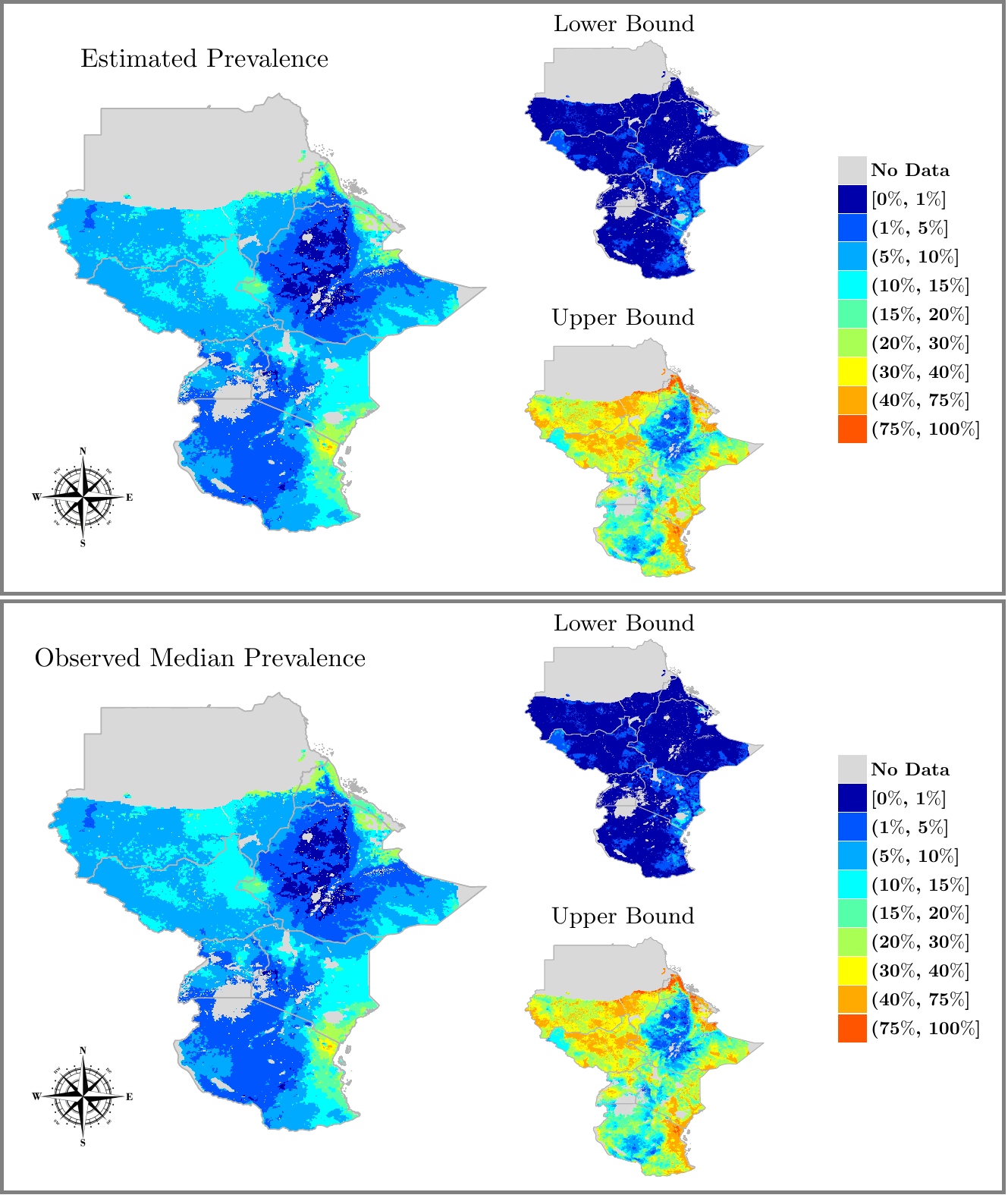}
\caption{Accuracy assessment of our method. The true distribution of LF prevalence (lower panel) is compared to the estimated prevalence distribution (upper panel) using our proposed methodology predicted at $5\times5$ km resolution. Point estimates along with lower (2.5\%) and upper (97.5\%) percentiles are presented.}
\label{fig:ObservedVsEstimated}
\end{figure}

In this section, the Bayesian approach presented in Section \ref{sec:Methods} was applied on the LF data. Firstly, we assessed the accuracy of the method, defined as the ability of the transmission model weighted simulations to reproduce the pre-control (baseline) geostatistical map, by comparing the observed and the estimated distribution of the baseline (equilibrium) mf prevalences at each pixel. Figure \ref{fig:ObservedVsEstimated} illustrates the median map (with 2.5 and 97.5 percentiles), along with the corresponding maps of the observed data. Overall, the results show that the maps are almost identical, indicating that the method is able to reproduce the distribution of the observed baseline prevalence in each pixel. In addition, in the left panel of Figure \ref{fig:PopandESS} of the Appendix we compared the estimated number of people per pixel with the observed value, which were in close agreement indicating that the proposed method accurately reproduced the number of people in each pixel. In the right panel of Figure \ref{fig:PopandESS} of the Appendix, we examined the ESS per pixel, which represents the effective number of simulations per pixel and is a measure of how well the method performs. We observed that the pixels with high prevalence (which may require a change of intervention strategy) have high ESS.

\begin{figure}[h!]
\centering
	\includegraphics[scale=0.68]{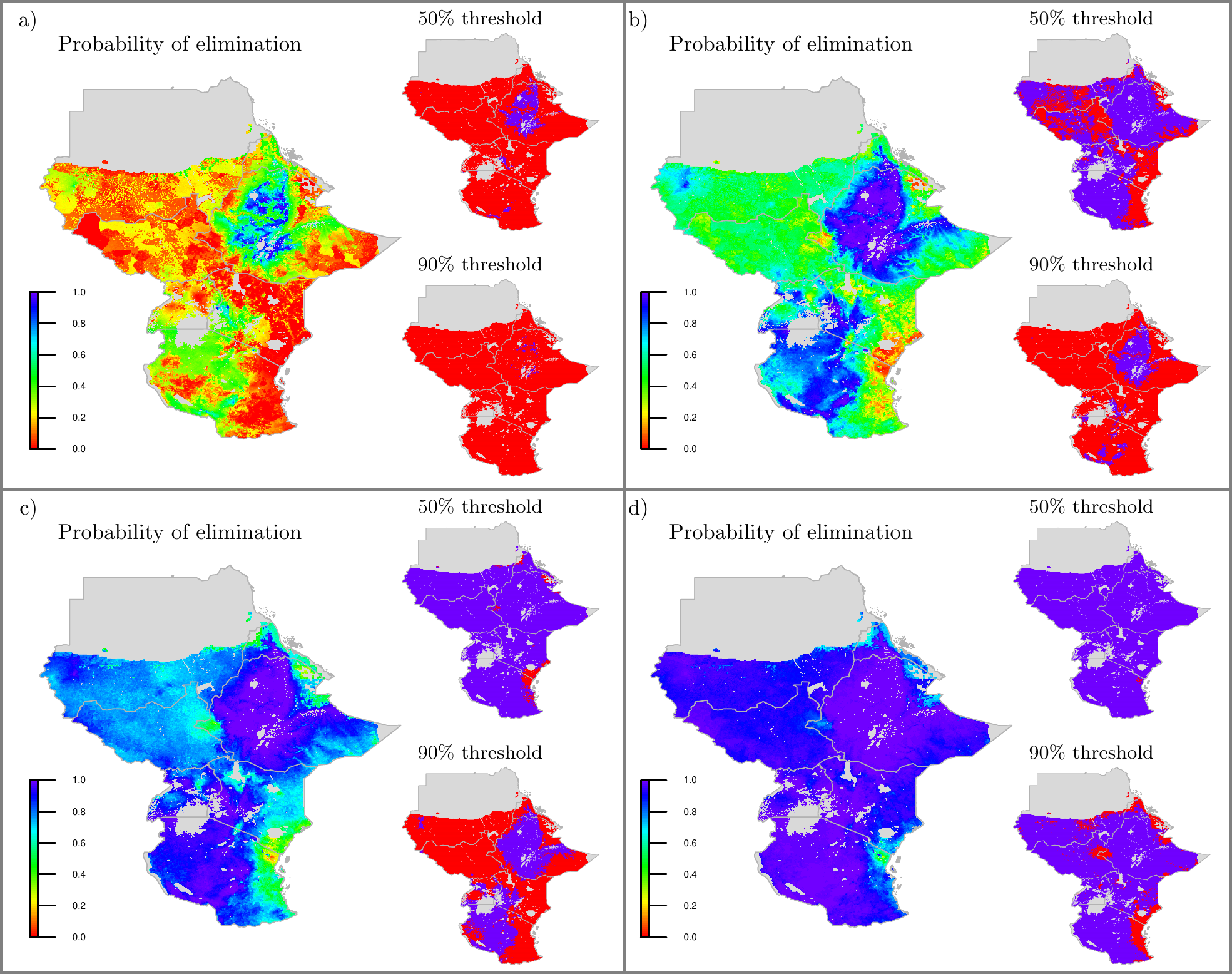}
\caption{Probability of less than 1\% prevalence after 5 years under: a) no intervention; annual MDA with coverage of b) 65\%; c) 80\% and d) biannual MDA at 65\% coverage predicted at $5\times5$ km resolution, for Ethiopia, Sudan, South Sudan, Eritrea, Kenya, Tanzania and Uganda.
Right of panels: Pixels that achieve elimination (blue) or do not achieve elimination (red) using different probability thresholds.}
\label{fig:PorbElimYear5}
\end{figure}

\begin{figure}[h!]
\centering
	\includegraphics[scale=1]{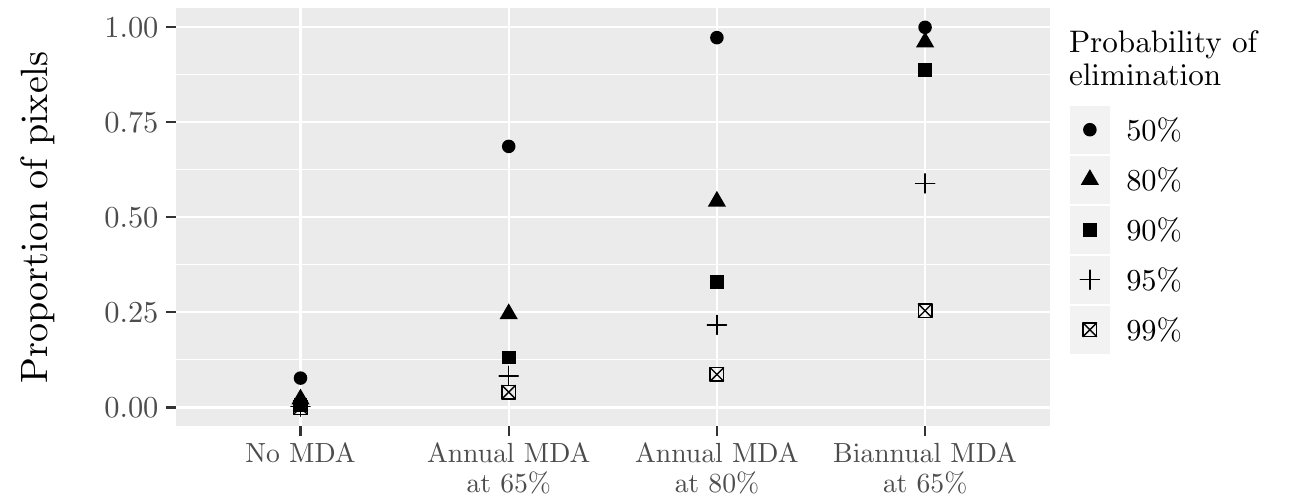}
\caption{Proportion of pixels with prevalence less than 1\% using different probability thresholds after 5 years under four intervention strategies. }
\label{fig:ProportionEliminated}
\end{figure}

Secondly, the methodology was applied to evaluate the impact of different intervention programs for LF in East Africa. In particular, four treatment scenarios were simulated: no interventions; the standard 65\% coverage annual MDA (aMDA); 80\% coverage aMDA; or biannual MDA (bMDA) at 65\% coverage, in order to investigate how these affect the probability of elimination after 5 years (Figure \ref{fig:PorbElimYear5}). 
Adopting a prevalence of less than 1\% as the threshold set by WHO as a global target for determining LF transmission elimination, our analysis predicted that the recommended strategy of 5 rounds of aMDA at 65\% is not enough for eliminating the disease in all pixels, with probability of elimination above 90\% only for 13\% of the pixels (see also Figure \ref{fig:ProportionEliminated}). 
Moreover, when more intensive treatments were implemented, i.e. more frequent MDA or higher coverage, the probability of elimination significantly increased compared to aMDA programme at 65\%. In particular, bMDA at 65\% coverage was the most effective of all strategies considered and was able to achieve elimination in 88\% of the pixels, with at least 90\% probability.
However, the proportion of pixels which achieved elimination after 5 years reduced to 59\% and  25\% when the probability threshold was increased to 95\% and 99\%, respectively, illustrating that the policy is sensitive to uncertainty. 

Finally, predictions of mf prevalence for the first and fifth year of intervention were summarized by calculating the estimated prevalence at each pixel, together with the 2.5\textsuperscript{th} and 97.5\textsuperscript{th} percentile in Figures \ref{fig:PrevYear1} and \ref{fig:PrevYear5} of the Appendix, for each of the four scenarios. Very similar observations were made on the predictions of mf prevalence for the first 5 years of intervention. 

\section{Discussion}
\label{sec:Discussion}

This study highlights the value of integrating geostatistical prevalence maps and transmission models for providing predictions on the impact of interventions aiming to eliminate transmission at a local scale. The main contribution is the development of new statistical tools through which existing research in mapping and predictive modelling are combined in a computationally efficient and flexible way which correctly accounts for uncertainty in these different techniques. Although we focus and apply our methodology on LF transmission, it can be applied to other infectious diseases.

We have shown that the current strategy of 5 annual rounds of MDA at 65\% coverage will not be sufficient to eliminate disease in most areas. We also found that a change in the current MDA strategy, such as increasing the coverage and frequency of MDA, will be required if LF elimination is to be accelerated in East Africa. This suggests that it may be necessary to employ different enhanced intervention plans at a fine scale, according to the characteristics of each area, in order to achieve the WHO elimination targets.

However, for the results presented here we assumed that no interventions have been applied in East Africa prior to the prevalence survey. While this assumption is correct for most areas, MDA programs began to be implemented in a few districts of Africa since 2000 and in many more districts thereafter. Therefore, one of the next steps will be to account for previous MDA programmes as spatio-temporal covariate information in the transmission model. Apart from MDA, insecticide treated bednets have been used in some countries (data can be extracted from the Malaria Atlas Project), the use of which has been shown to be an effective additional measure for control of the disease  \citep{Bockarie:2009}. Integrating geostatistical maps with transmission models with these additional covariates is more complicated as the simulations must include the appropriate historical interventions. 

An additional challenge is the gap between reported and true coverage with an MDA. Where there are parasitological data against which to test the expected and achieved impact of reported coverages, they have been shown to be unreliable \citep{Budge:2016}. This will pose a particular challenge to interpreting historic coverage and a challenge in communicating future projections. This work represents our initial framework and future research will be required to extend the methodology to capture these more complex settings.

A limitation of our statistical approach is that it doesn't capture the spatial correlations in the predictions, since each pixel is weighted independently to produce a marginal posterior for each pixel. This approach means that we lose the spatial autocorrelation that was captured in the original geostatistical model and, furthermore, that there is no way for nearby pixels to interact during the simulations, for example, to account for movements of humans or vectors. A more sophisticated approach would be to use the spatial autocorrelations from the geostatistical model, alongside any available movement or connectivity data, to define a transmission kernel that describes spatial spread. This kernel could be used within a single meta-population model describing the transmission dynamics across the whole map. At present, such an approach would be computational infeasible at the country scale, but may become possible in future through improvements in methodology and advances in parallelisation and cloud computing.

\section*{Acknowledgments}

The authors gratefully acknowledge funding of the NTD Modeling Consortium by the Bill and Melinda Gates Foundation [OPP1152057, OPP1053230, OPP1156227, OPP1186851]. The views, opinions, assumptions or any other information set out in this article should not be attributed to the Bill and Melinda Gates Foundation or any person connected with the Bill and Melinda Gates Foundation. The authors thank Rachel L.~Pullan and Jorge Cano for sharing the geostatistical maps of LF prevalence and Michael A.~Irvine and Paul Brown for improving the model code. We also thank Dr.~Nick Golding for providing helpful comments on our manuscript.

\section*{Appendix}
\renewcommand\thefigure{\thesection.\arabic{figure}} 
\renewcommand\thetable{\thesection.\arabic{table}} 
In the appendix, we provide further details on the methodology and the transmission model as well as additional plots and results of the real data analysis.

\appendix

\section{Background information on the methodology}
\setcounter{table}{0}
\setcounter{figure}{0}
\subsection{Proof of lemma on the change of measure}\label{proof}

\begin{proof} 
First we need to change the variables of the prior from $\btheta$ to the transformed variables $(p,\bq)$. This results in a transformed prior density $\pi'(p,\bq)=\pi(\bphi^{-1}(p,\bq))|J^{-1}|,$ where $J^{-1}$ is the determinant of the Jacobian matrix of partial derivatives of $\bphi^{-1}$. Next, we wish to apply the Radon-Nikodym derivative formula \cite{} to change the measure over prevalences from $g(p)$ to $f(p)$, whilst keeping the measure over $\bq$ unchanged. More formally, we rewrite $\pi'(p,\bq)=\pi'(p)\pi'(p|\bq)=g(p)\pi'(p|\bq)$ and change the whole measure from $g(p)\pi'(p|\bq)$ to $f(p)\pi'(p|\bq)$. This yields the new measure in terms of the transformed variables $h'(p,\bq)=\frac{f(p)\pi'(p|\bq)}{g(p)\pi'(p|\bq)}\pi'(p,\bq)=\frac{f(p)}{g(p)}\pi'(p,\bq)$. The final step is to transform back to the original variables $\btheta$.
\begin{align*}
h(\btheta)&=h'(\phi(\btheta))|J|\\
&=\frac{f(p(\btheta))}{g(p(\btheta))}\pi'(\phi(\btheta))|J|\\
&=\frac{f(p(\btheta))}{g(p(\btheta))}\pi(\btheta).
\end{align*}
\end{proof}

\subsection{Implementation considerations from the absolute continuity condition}\label{sec:SM_note2}

The absolute continuity condition in Lemma \ref{lemma}, that $f$ must be absolutely continuous with respect to $g$, means that whenever $g(p)=0$ then we must also have $f(p)=0$. In other words, when the prior probability of a prevalence is zero then the map measure of prevalence must also be zero. This has important considerations for implementing our method. For example, we find that our prior for the transmission model parameters is unlikely to produce any simulations with a prevalence of above 85\%. However, prevalences this high do appear in the tails of the posterior distribution in the geostatistical model, especially when the amount of uncertainty is high. When the weighting is applied, there are no simulations to capture the high prevalence region, which leads to an underestimation of the mean prevalence, for example.

There are several potential approaches to ameliorate this. The simplest would be to mask out areas of the map where this occurs. However, there may be large areas of the map that have only a very small probability of producing a prevalence above 85\% according to the geostatistical model. A second approach, which we could take if we believed that prevalences above 85\% were a priori impossible, would be to remove any samples from the geostatistical map that went above 85\%. This is effectively the same as applying the weighting naively, and leads to the mean and median of the prevalence distribution of the simulations being lower than the corresponding statistics from the geostatistical model. A final more sophisticated approach could be to shift the weight from high prevalences in the geostatistical map to the nearest prevalences available in the simulations. For example, by using the histogram-based empirical Radon-Nikodym derivative with a wide bin capturing all prevalences above 75\%. Although this reduces the size of the underestimation, it substantially reduces the effective sample size, as a small number of simulations get very high weight. 

\subsection{Minimum discrepancy-based empirical Radon-Nikodym derivative}
\label{sec:SM_Minimum}

\begin{figure}[h!]
\centering
	\includegraphics[scale=0.87]{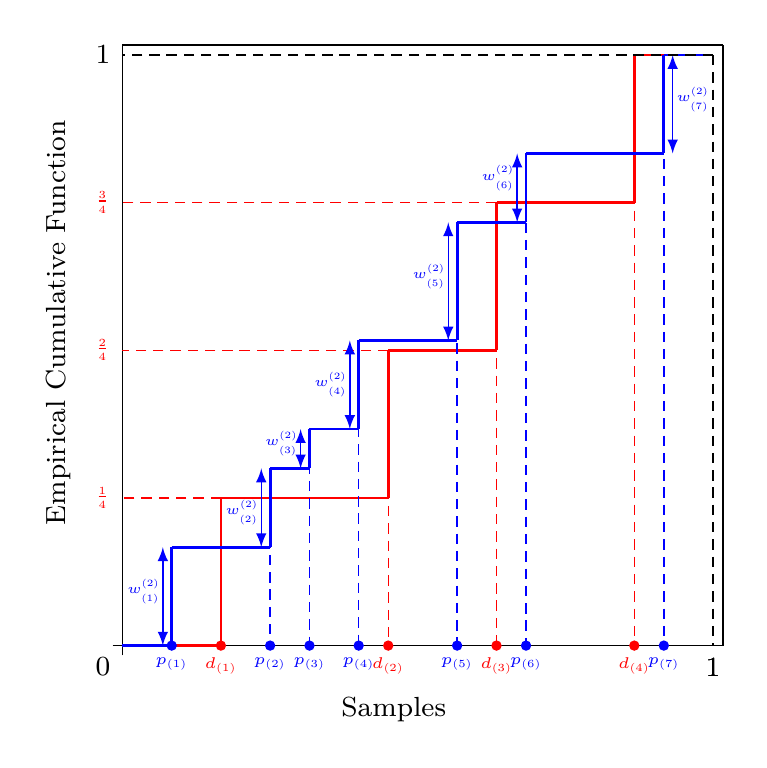} 
		\includegraphics[scale=0.87]{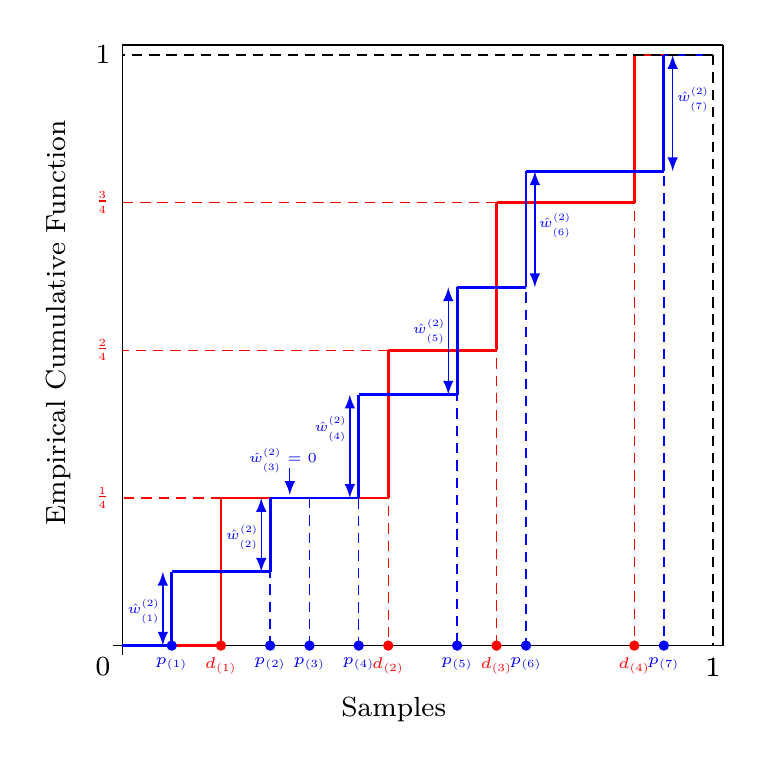}
\caption{An example of the empirical cumulative distribution functions of the posterior prevalences (red) and the weighted simulated prevalences (blue). The empirical cdf of the weighted simulated prevalence is shown in the left panel with arbitrary weights and in the right panel using the optimal weights that achieve the minimum distance between the two cdfs, as described in Section \ref{sec:SM_Minimum}.}
\label{fig:ExampleCdfs}
\end{figure}

For each pixel, we would like to minimise the following distance between the empirical cumulative distribution functions (cdfs) of the posterior prevalences and the weighted simulated prevalences:
$$\int_0^1 \left(F(x|\bd)-H(x|\bw^{(2)})\right)^2\:\mathrm{d}x,$$
with respect to the weights $\bw^{(2)}$. First, the posterior and simulated prevalence samples are sorted in ascending order, such that $d_{(1)} \leq d_{(2)}  \leq \ldots  \leq d_{(M-1)} \leq d_{(M)}$ and $p_{(1)} \leq p_{(2)} \leq \ldots \leq p_{(J-1)} \leq p_{(J)}$. Similarly, we introduce the notation $w^{(2)}_{(j)}$ for the weights of simulation $j$ which produces prevalence $p_{(j)}$, for $j = 1, 2, \ldots, J$. For clarity of notation, an example of two empirical cdfs is shown in the left panel of Figure \ref{fig:ExampleCdfs}.  

Starting with $j=1$, we have that:
\begin{align}
 \frac{\partial }{\partial w^{(2)}_{(1)}} &\int_{p_{(1)}}^{p_{(2)}} \left(F(x|\bd)-H(x|\bw^{(2)})\right)^2\:\mathrm{d}x =  \frac{\partial }{\partial w^{(2)}_{(1)}} \int_{p_{(1)}}^{p_{(2)}} \left(F(x|\bd)-w^{(2)}_{(1)}\right)^2\:\mathrm{d}x \nonumber\\
 &= -2 \left[ \int_{p_{(1)}}^{p_{(2)}} \left(F(x|\bd)-w^{(2)}_{(1)}\right)\:\mathrm{d}x\right] \nonumber \\
 &= -2 \left[ \int_{p_{(1)}}^{p_{(2)}} F(x|\bd)\:\mathrm{d}x - w^{(2)}_{(1)}\left(p_{(2)} - p_{(1)}\right) \right].
 \label{eq:j=1}
\end{align}

At the minimum Equation \eqref{eq:j=1} is equal to 0, therefore:
\begin{equation*}
 \hat{w}^{(2)}_{(1)} =  \frac{\displaystyle \int_{p_{(1)}}^{p_{(2)}} F(x|\bd)\:\mathrm{d}x}{p_{(2)} - p_{(1)} }.   
\end{equation*}

Similarly, we differentiate the distance between the two cdfs with respect to $w^{(2)}_{(j)}$, for $j = 2, 3, \ldots, J-1$ and by setting it equal to 0,
\begin{align*}
 \frac{\partial }{\partial w^{(2)}_{(j)}} &\int_{p_{(j)}}^{p_{(j+1)}} \left(F(x|\bd)- \sum_{k=1}^{j-1} \hat{w}^{(2)}_{(k)} - w^{(2)}_{(j)} \right)^2\:\mathrm{d}x \\
 &= -2 \left[ \int_{p_{(j)}}^{p_{(j+1)}} F(x|\bd)\:\mathrm{d}x - \left(p_{(j+1)} - p_{(j)}\right) \left( \sum_{k=1}^{j-1} \hat{w}^{(2)}_{(k)} + w^{(2)}_{(j)} \right)  \right] = 0,
\end{align*}
we obtain:
\begin{equation*}
\hat{w}^{(2)}_{(j)} =  \frac{\displaystyle \int_{p_{(j)}}^{p_{(j+1)}} F(x|\bd)\:\mathrm{d}x}{p_{(J+1)} - p_{(J)} } - \sum_{k=1}^{j-1} \hat{w}^{(2)}_{(k)}.  
\end{equation*}

\noindent Finally, to ensure that the weights sum to one, for $j=J$ we have that:
\begin{equation*}
 \hat{w}^{(2)}_{(J)} =  1 - \sum_{k=1}^{J-1} \hat{w}^{(2)}_{(k)}
\end{equation*}

\noindent As an example, in the right panel of Figure \ref{fig:ExampleCdfs} we provide the optimal weights $\hat{w}^{(2)}_{(j)}$, i.e.~the ones with the minimum distance between the two empirical cdfs.

\section{Toy example}
\label{sec:Toyexample}
\setcounter{table}{0}
\setcounter{figure}{0}

\begin{figure}[h!]
\centering
	\includegraphics[scale=1]{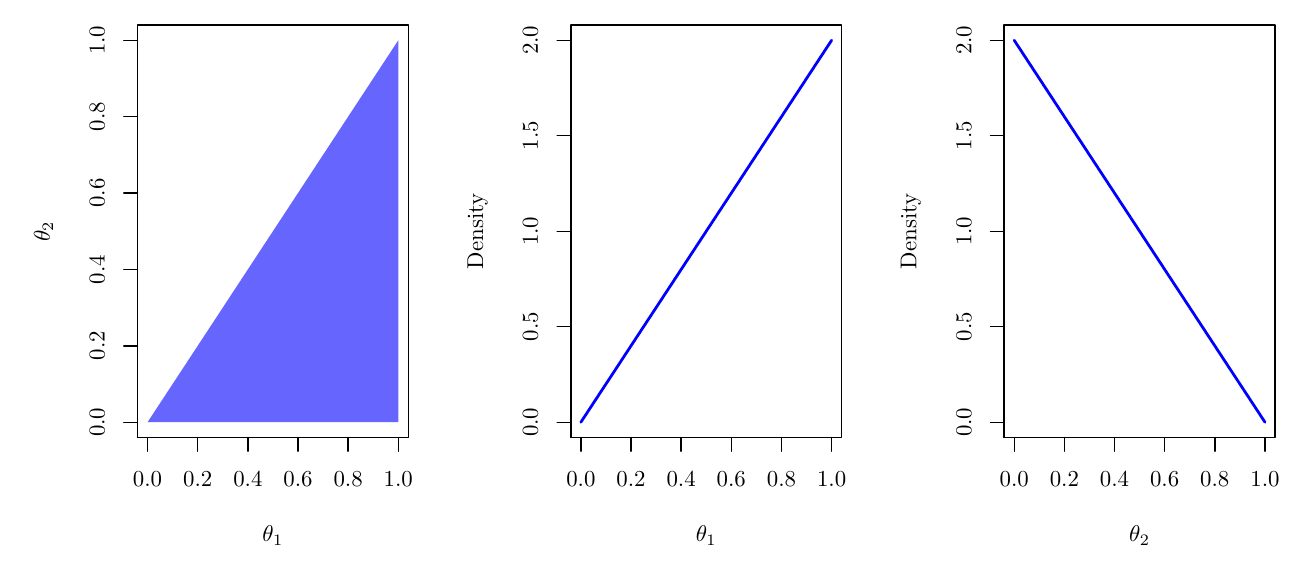}
\caption{Prior support and marginal densities for the parameters of the toy example in Section \ref{sec:Toyexample}.}
\label{fig:toyPrior}
\end{figure}

Suppose that the prior distribution is $\pi(\theta_1,\theta_2)=2$ if $0<\theta_2<\theta_1<1$ and zero otherwise. The prior support and marginal densities are shown in Figure \ref{fig:toyPrior}. For simplicity, assume that the transmission model has equilibrium prevalence given by $p(\theta_1,\theta_2)=\theta_1$ so that the induced prior over prevalences is the marginal for $\theta_1$, ie.~$g(p)=2p$ for $0<p<1$. Further, suppose that we are given a pixel with prevalence measure $f(p)=2(1-p)$ for $0<p<1$. This challenging example allows us to assess how the methodology performs when there are few simulations with low weights in the area of high posterior probability close to $p=0$. 

By the Lemma \ref{lemma} of the main text, the new measure over the parameter space is given by
\begin{align*}
h(\theta_1,\theta_2)&=\frac{f(p(\theta_1,\theta_2))}{g(p(\theta_1,\theta_2))}\pi(\theta_1,\theta_2)\\
&=\frac{f(\theta_1)}{g(\theta_1)}\times 2\\
&=\frac{2(1-\theta_1)}{2\theta_1}\times 2\\
&=\frac{2(1-\theta_1)}{\theta_1} \qquad\text{for $0<\theta_2<\theta_1<1$.}
\end{align*}

First, we can verify that we get the correct marginal for $p=\theta_1$.
\begin{align*}
f(p)&=\int h(p,\theta_2) \:\mathrm{d}\theta_2\\
&=\int_0^p 2\frac{(1-p)}{p} \:\mathrm{d}\theta_2\\
&=\left[\theta_2 \frac{2(1-p)}{p}\right]_{\theta_2=p}\\
&=2(1-p) \qquad\text{for $0<p<1$,}
\end{align*}
as required.

Second, we can determine the new marginal density for $\theta_2$.
\begin{align}
h(\theta_2)&=\int h(\theta_1,\theta_2) \:\mathrm{d}\theta_1 \nonumber\\
&=\int_{\theta_2}^1 \frac{2(1-\theta_1)}{\theta_1} \:\mathrm{d}\theta_1 \nonumber\\
&=2\left[ \log(\theta_1)-\theta_1\right]_{\theta_1=\theta_2}^1 \nonumber \\
&=2(\theta_2-\log(\theta_2)-1)\qquad\text{for $0<\theta_2<1$.}
\label{eq:PosteriorTheta2}
\end{align}

\subsection{Simulations}

In this section, we perform a series of simulations to assess the accuracy and efficiency of the proposed method for recovering the distribution of pixel prevalence. Particular focus is given on how the method is affected as we vary the value of $\delta$, the proposal distribution $q(\btheta)$, and the empirical estimate of the Radon-Nikodym derivative.

\subsubsection{Sensitivity analysis: Value of $\delta$.}\label{toydelta}

We first investigated the performance of the proposed method as a function of $\delta$, where the observed pixel and simulated prevalence data are obtained from the toy model described in Section \ref{sec:Toyexample}. In particular, we generated $M=2\,000$ pixel prevalence samples from Beta(1,2). We then generated $J=2\,000$ samples from the joint prior distribution of parameters $\theta_1$ and $\theta_2$; first we drew $\theta_1 \sim $ Beta(2,1) and then $\theta_2 \mid \theta_1 \sim \theta_1 \times$Beta(1,1). Therefore, the obtained simulated prevalence samples are draws from Beta(2,1), which is the marginal prior for $\theta_1$. We considered values for $\delta$ from 0.001 through to 0.1, increasing by 0.001 each time.

\begin{figure}[h!]
\centering
\includegraphics[scale=1]{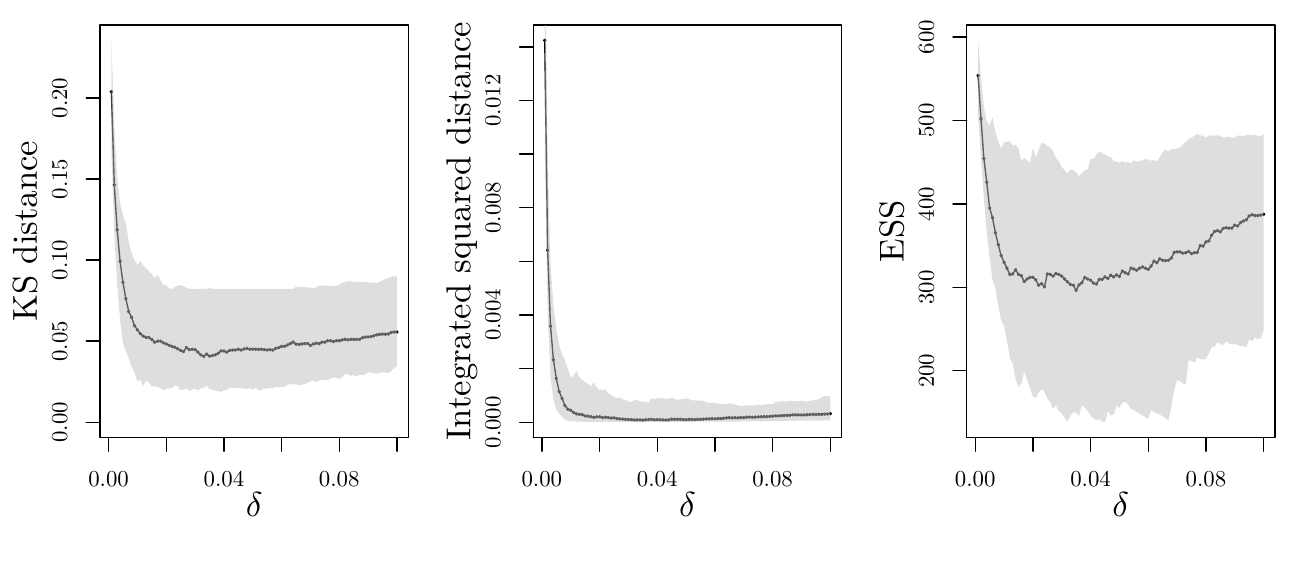}
\caption{Sensitivity to $\delta$ in the example where the parameters are drawn from their prior distribution. The left panel shows the Kolmogorov-Smirnov (KS), the middle panel the integrated squared distance and the right panel the ESS for $\delta = 0.001, 0.002, \ldots, 0.1$. Shaded areas correspond to the 95\% credible interval.}
\label{fig:KSESSToyPrior}
\end{figure}

Accuracy was assessed by computing the Kolmogorov–Smirnov (KS) distance which is defined as the largest vertical distance between the two empirical cumulative distribution functions of the pixel prevalence and the weighted simulated prevalence. We also consider an additional measure of quantifying the distance between the two cdfs, termed as integrated squared distance and given by $\int_0^1 \left(F(x|\bd)-H(x|\bw^{(2)})\right)^2\:\mathrm{d}x$. The calculation of the two distances is repeated 100 times for each value of $\delta$, using new pixel and simulated prevalence samples each time, in order to prevent biases occurring due to the simulating procedure. Results are shown in Figure \ref{fig:KSESSToyPrior}. We see that the performance of our method is affected by the value of $\delta$. In the left panel of Figure \ref{fig:KSESSToyPrior} we show the median KS distance along with the 95\% credible interval, over the 100 realisations for each $\delta$. Overall, the accuracy of the algorithm increases as $\delta$ grows from 0.001 to 0.033, where the median KS distance reaches its minimum value, and for $\delta$ higher than 0.033 a slight decrease is observed. The integrated squared distance, shown in the middle panel of  Figure \ref{fig:KSESSToyPrior}, provides identical conclusions. Therefore, from now on we use the integrated squared difference to assess the accuracy of the method. An opposite pattern is observed in the efficiency of the method, as can be seen in the right panel of Figure \ref{fig:KSESSToyPrior} where we show the effective sample size (ESS) as a function of $\delta$. Note that despite having the highest ESS for $\delta = 0.001$, the method has the lowest accuracy as indicated by the highest distance between the two empirical distributions. This could be attributed to the fact that there are very few simulations in each estimate of density $g$ (the denominator in Equation \ref{eq:Methods1}).

\begin{figure}[h!]
\centering
\includegraphics[scale=1]{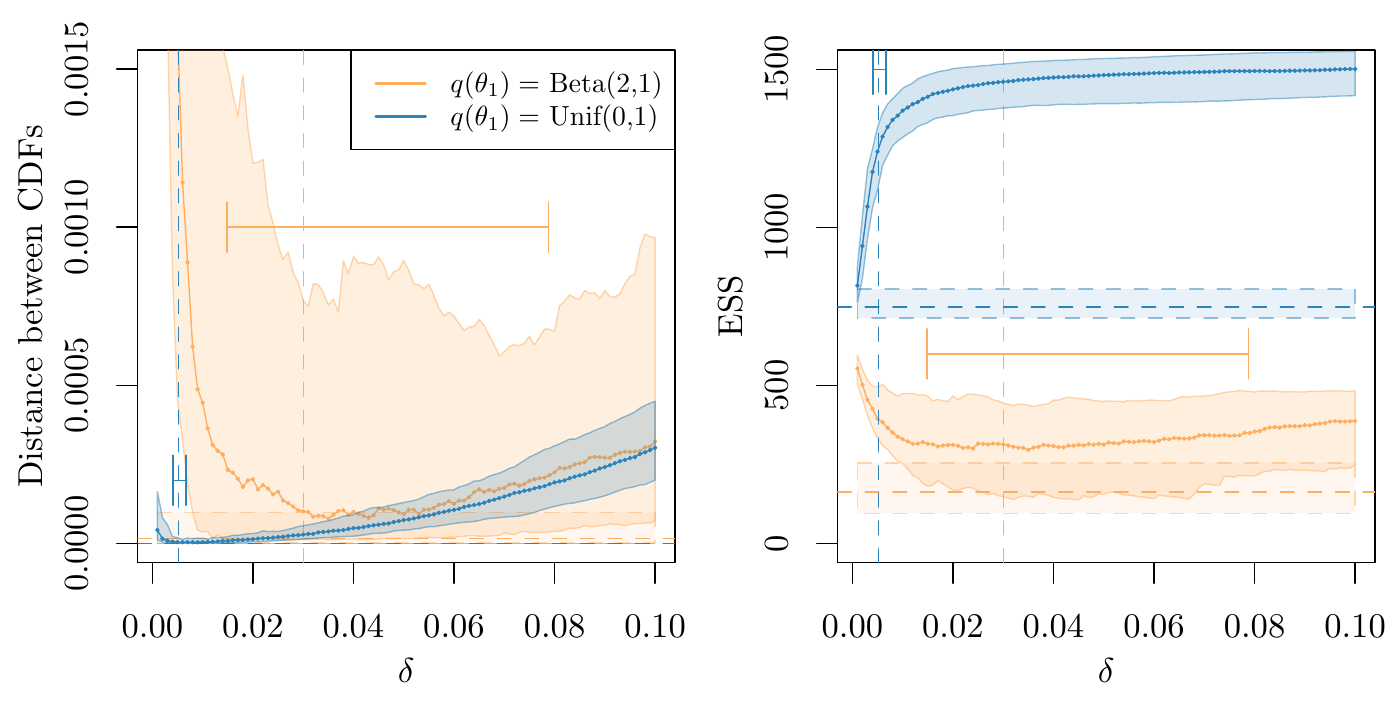}
\caption{Integrated squared distance between the two cumulative distribution functions (cdfs) (left panel) and effective sample size (ESS) (right panel) obtained from 100 simulated datasets, under different values of $\delta$ and choice of proposal distribution for parameter $\theta_1$. Orange solid line represents a prevalence proposal distribution equal to marginal prior, i.e.~Beta(2,1), whereas blue line corresponds to a $U(0,1)$ proposal. In both cases, the pixel prevalences were drawn from a Beta(1,2) distribution. Dashed horizontal lines correspond to the minimum possible distance (left panel) and its associated ESS (right panel). Shaded areas correspond to the 95\% credible intervals for each estimate. Dashed vertical lines represent the suggested value of $\delta$ for each scenario of the proposal distribution considered, along with their upper and lower limits (error bars). }
\label{fig:Distance_ESS_Toy}
\end{figure}

This motivated us to propose an appropriate value of $\delta$ depending on the simulated prevalence samples. More specifically, let $\delta_{k,j} = 2 \mid p_k - p_j \mid$, for $j, k = 1, 2, \ldots, J$. For each $k$, let $\delta_{k,(j)}$ to denote the sorted values with respect to $j$, such that $\delta_{k,(1)} \leq \delta_{k,(2)} \leq \ldots \leq \delta_{k,(J)}$. Note that $\delta_{k,(1)}$ is always equal to zero since for $k=j$, $\delta_{k,k} = 0$. Finally, the suggested value of $\delta$, denoted by $\tilde{\delta}$, is given by $\tilde{\delta} = \max \left( \delta_{1,(3)}, \delta_{2,(3)}, \ldots, \delta_{J,(3)} \right)$, ensuring that at least 3 simulations were included in each estimate of $g$. In Figure \ref{fig:Distance_ESS_Toy}, we report the median $\tilde{\delta}$ (orange vertical line) along with the 95\% credible interval, as obtained from the 100 simulations. Overall, the suggested value of $\delta$ achieves a good balance between efficiency and accuracy. In addition, the estimated weighted prevalence distribution of the suggested value of $\delta$ is provided in the right panel of Figure \ref{fig:BestDeltaToyPrior} in the main text and shows that the method correctly reproduces the histogram of the true pixel prevalence (left panel), except maybe for prevalence values close to 0 where the histogram is a bit noisy. This is justified by the fact that there are only few simulations that have prevalence values between 0 and 0.02, as illustrated in the middle panel.

\subsubsection{Sensitivity analysis: Proposal distribution.}

In this section, we carry out a sensitivity analysis to assess the effect that the proposal distribution of the parameters has on the performance of the method. More specifically, instead of drawing $\theta_1$ from its prior distribution Beta(2,1), we consider a uniform proposal $U(0,1)$. This change results to a uniform distribution over the simulated prevalences. As before, we assess the performance of the method for different values of $\delta$, shown in Figure \ref{fig:Distance_ESS_Toy}. Overall, we conclude that the performance of the method improves substantially when we move from the prior to the uniform proposal distribution over the prevalences. This is because there are no areas with strong posterior probability that have few simulations from the proposal. In particular, the median integrated squared distance between the two cdfs (left panel) appears to be much lower in the latter compared to the former, and it is also associated with lower variability of the estimate. In addition, using a
uniform proposal distribution leads to substantial increase in ESS (right panel), with the median ESS being at least 1.5 times higher for values of $\delta$ close to 0.001 and up to 3.5 times for larger values of $\delta$. For reference, we also display the minimum discrepancy-based empirical Radon-Nikodym derivative, as described in Section \ref{sec:SM_Minimum}, and its associated ESS. Overall, we come close to reach the minimum possible integrated squared distance between the two cdfs, with larger ESS. 

\begin{figure}[h!]
\centering
\includegraphics[scale=1]{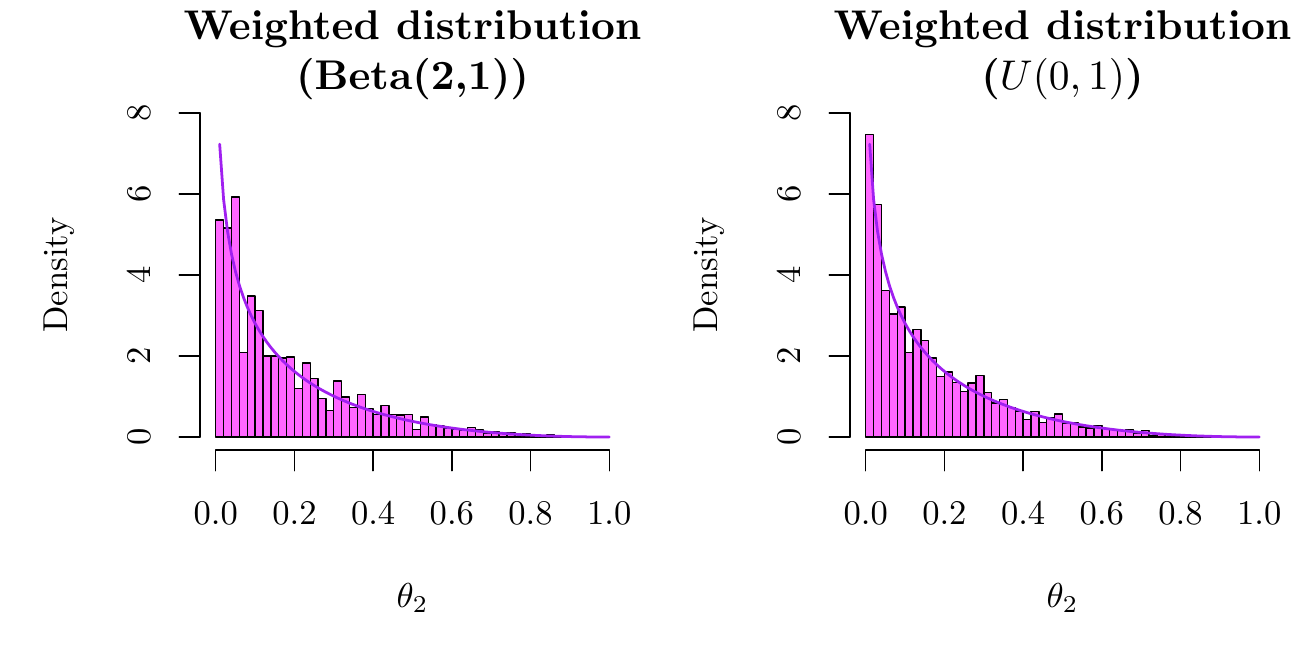}
\caption{The estimated weighted posterior distribution of $\theta_2$ is compared to the target density (purple line), under different proposal distributions for the prevalence: Beta(2,1) (left panel) and $U(0,1)$ (right panel).}
\label{fig:PosteriorTheta2}
\end{figure}

To further assess the accuracy of the method we compared the weighted posterior distribution of $\theta_2$ with its true density, calculated in Equation \eqref{eq:PosteriorTheta2}. Results are shown in Figure \ref{fig:PosteriorTheta2} for each choice of the proposal distribution. The simulation analysis illustrates that both proposals perform well in terms of recovering the target density of $\theta_2$. Nevertheless, using a uniform proposal for the prevalence leads to a large improvement in the estimate.

\subsubsection{Sensitivity analysis: Empirical estimate of the Radon-Nikodym derivative.} 

In order to study the influence of the empirical Radon-Nikodym derivative (ERND) on the method performance, we repeated our simulation analysis for three different choices: (a) the proposed
empirical estimate described in Section \ref{step2} referred as the distance-based derivative, (b) the histogram-based derivative in Section \ref{sec:histogram} and (c) the discrepancy-based derivative using the integrated squared distance (for more details see Section \ref{sec:discrepancy} of the manuscript). Results are given in Table \ref{tab:ERND}. Note that the proposed distance-based derivative is based on using the prevalences within $\tilde{\delta}/2$ of $p_j$ and the histogram-based derivative is calculated by splitting the $[0, 1]$ interval into 100 equal bins.

\begin{figure}[h!]
\centering
\includegraphics[scale=1]{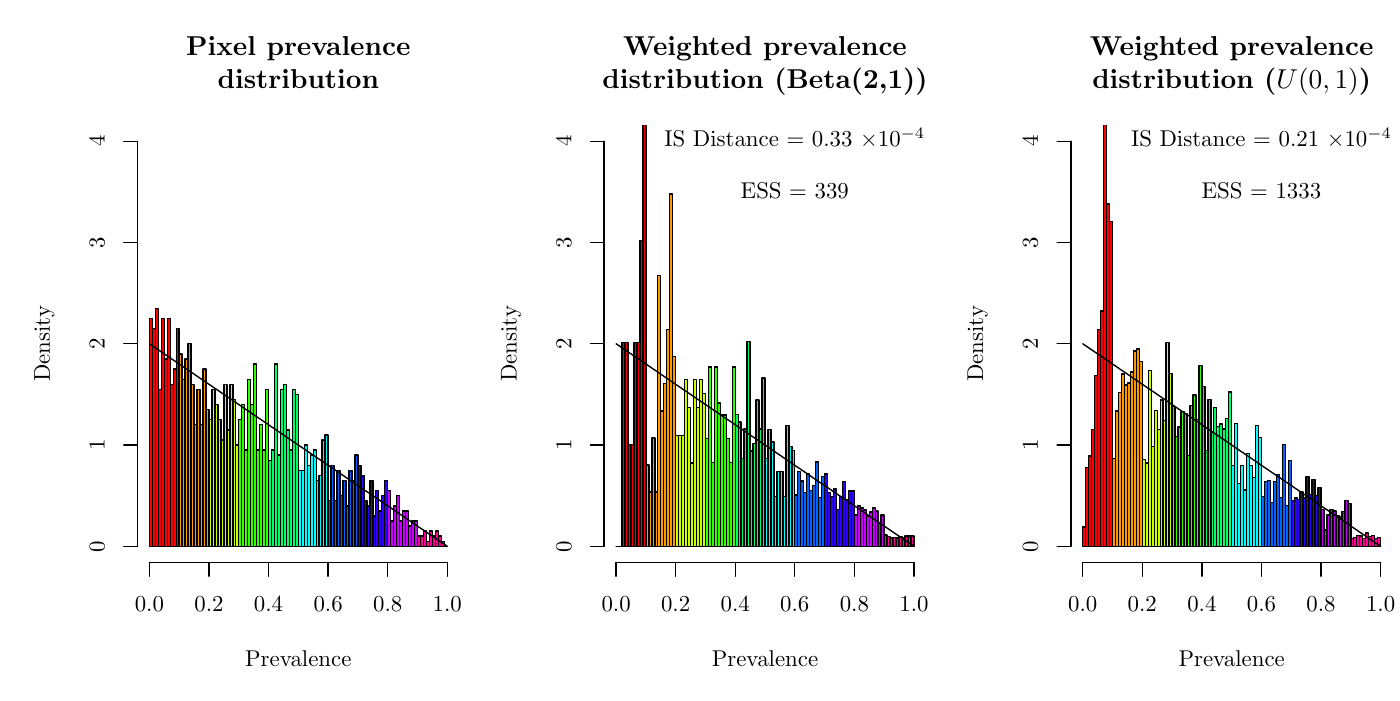}
\caption{The estimated weighted prevalence distributions obtained using the histogram-based empirical Radon-Nikodym derivative for 10 bins using a Beta(2,1) (middle panel) and $U(0,1)$ proposal (right panel) is compared to the true pixel prevalence distribution (left panel). Different colours correspond to prevalences in different bins. For reference we also provide the integrated squared (IS) distance and the ESS.} 
\label{fig:HistogramBasednbins}
\end{figure}

As expected, the discrepancy-based ERND is optimal in terms of the distance between the two cdfs, but results in much lower ESS compared to the alternative derivatives. When there are few simulations of the proposal in the area of high posterior probability close to zero prevalence, the distance-based ERND performs better compared to the histogram-based ERND in terms of both integrated squared distance and ESS. When we move to a uniform simulated prevalence distribution, we find that histogram-based derivative outperforms the distance-based derivative, and is the one that scores highest in terms of ESS, followed in order by distance-based and discrepancy-based derivative. However, its performance depends on the choice of bins, since the relative weightings within each bin are unchanged. For example, Figure \ref{fig:HistogramBasednbins} uses just 10 bins to illustrate that the distribution within each bin does not reflect the target distribution.
Finally, we conclude that the performance of all the derivatives is greatly improved when we change the proposal from the prior to the uniform distribution over the prevalences.

\begin{table}[h!]
\centering
\begin{tabular}{|c c c c|}
\hline   \multicolumn{1}{|c}{Proposal} & \multicolumn{1}{c}{\multirow{2}{*}{  ERND} }   & \multicolumn{1}{c}{Integrated squared} &  \multicolumn{1}{c|}{\multirow{2}{*}{  ESS} } \\
 \multicolumn{1}{|c}{distribution} &   & \multicolumn{1}{c}{distance ($\times 1000$)} &
  \\
  \hline \hline 
\multirow{3}{*}{Beta(2,1)} & Distance & 0.23734 (0.01742, 1.03220) & 338  (222, 473) \\[8pt]
 & Histogram &  0.46996 (0.02757, 1.56949)
 & 335  (252, 453) \\[8pt]
  &Discrepancy& 0.01647 (0.00335, 0.09844) & 164 (96, 255) \\[8pt] \hline 
  \multirow{3}{*}{$U$(0,1)} & Distance & 0.00408 (0.00157, 0.01799) & 1248 (1161, 1330)\\[8pt] 
 & Histogram & 0.00214 (0.00175, 0.00292) & 1347 (1253, 1429) \\[8pt]
  &Discrepancy & 0.00025 (0.00021, 0.00029) & 748 (713, 805)
   \\[8pt] 
\hline  \end{tabular}
\caption{Integrated squared distance between the two empirical cdfs (multiplied by 1000) and ESS, averaged over 100 replicates under three different empirical Radon-Nikodym derivatives (ERND) and different proposal distributions. The 95\% credible intervals are shown in parentheses.}
\label{tab:ERND}
\end{table}

\section{The mathematical model of LF transmission dynamics}
\label{sec:TRANSFIL}

\setcounter{table}{0}
\setcounter{figure}{0}

\subsection{TRANSFIL model description}

We employed the mathematical model of lymphatic filariasis (LF) transmission TRANSFIL \citep{Irvine:2015} to carry out the analysis in this paper. The model is a stochastic microsimulation of individuals with worm burden, microfilariae (mf) and other demographic parameters in relation to age and exposure to risk. 
Humans are modelled individually with their own burden of male and female worms denoted by $W_i^m$ and $W_i^f$, respectively. Mf concentration in the peripheral blood, denoted by $M_i$, is also modelled for each individual and according to the number of fertile female worms is increasing as well as decreasing at constant rate. The total mf density in the population contributes towards the current density of L3 larvae in the human-biting mosquito population.
Therefore, the model describes the dynamics of individual human, worm inside the host, mf inside the host and larvae inside the mosquito. A detailed description and mathematical formulation of the model are given in  \cite{Irvine:2015} and more recently in \cite{Smith:2017}, so here we provide a summary. 

\begin{description}
 \item[ Worm dynamics] \hfill \\ 
 For each individual $i$, both male and female worms are added according to their bite risk $b_i$ which is individually sampled from a gamma distribution with mean = 1 and shape parameter $k$. The rate at which an individual $i$ acquires an adult worm, either female or male, is described by the following expression:
 \begin{align*}
 0.5 \lambda b_i (V/H) \psi_1 \psi_2 s_2 h(a),
 \end{align*}
where $\lambda$ is the number of bites per mosquito, $V/H$ is the ratio of vectors to hosts, $\psi_1$ is the probability that an L3 larvae leaves the host during a biting event, $\psi_2$ is the probability that the L3 enters the host, $s_2$ is the proportion of L3 within the host that develop into adult worms and $h(a)$ is the age-dependent biting rate which increases with body size to saturate at age nine.  Finally, we assume that each worm has a constant death rate $\mu$.

  \item[ Mf dynamics] \hfill \\ 
  For an individual $i$, the dynamics of mf are given by the following expression:
   \begin{align*}
   \frac{\text{d}M_i}{\text{d}t} = \alpha W_i^f \mathbb{I}(  W_i^m>0) - \gamma M_i,
    \end{align*}
    with $\alpha$ being the production rate of mf per worm, $\gamma$ the constant death rate of mf and function $\mathbb{I}$ is one if there are male worms and zero if not. This expression follows from the fact that \textit{W. bancrofti} is completely polygamous and therefore the mf production rate depends  upon the number of female worms combined with the presence of at least one male in the human host.
     
  \item[ Larvae dynamics] \hfill \\
  Larvae development occurs when mf entered the mosquito during a blood meal form an infected host. There are two forms of this relationship depending on the genus of mosquito vectors as expressed below:
   \begin{align*}
   L(m) & = \kappa_{s1}\left( 1 - e^ {-r_1m/\kappa_{s1}} \right) \text{\quad for \textit{Culex}}, \\ 
      L(m) & = \kappa_{s2}\left( 1 - e^ {-r_2m/\kappa_{s2}} \right)^2 \text{\quad for \textit{Anopheles}}, 
    \end{align*}
    where $m$ is the concentration of mf per $20\mu$L taken during a blood meal and $r, \kappa_s$ are the parameters which are related to the functional form of the uptake curve \citep{Gambhir:2008}. The uptake of larvae is an average of mf concentration in the peripheral blood over all individuals  weighted by their bite-risk $b_i$ and is described by the following function:
       \begin{align*}
       \tilde{L} = \frac{\sum_i L(m_i) b_i}{\sum_i b_i},
    \end{align*}
giving the average number of larvae per mosquito. Using this expression we can calculate the averaged number of larvae taken up in the population $ \tilde{L}$ as follows:
      \begin{align*}
      \frac{\text{d}L}{\text{d}t} = \lambda g \tilde{L} - ( \sigma + \lambda \psi_1)L,
    \end{align*}
    where $\lambda$ is the number of bites per mosquito, $g$ is the proportion of mosquitoes which pick up infection when biting an infected host, $\sigma$ is the death rate of mosquitoes and $\psi_1$ is the proportion of L3 leaving the mosquito per bite.
 
Finally the equilibrium value for L3 in a mosquito is given by:
      \begin{align*}
      L^* = \frac{\lambda g \tilde{L} }{\sigma + \lambda \psi_1}.
    \end{align*}

  \item[Host dynamics] \hfill \\ 
Each human begins with zero infection and as mentioned before, has a bite-rate of exposure drawn from a gamma distribution with mean = 1 and shape parameter $k$. The shape parameter defines how aggregated bites are amongst individuals and consequently defines the aggregation of infection amongst individuals \citep{Irvine:2015}. Each individual has a rate of death $\tau$ that is assumed to be constant throughout an individual’s lifetime with a cut-off at age 100.
\end{description}

A list of the basic model parameter values is provided in Table \ref{tab:Parameters}, with specification of their source. In Table \ref{tab:ParametersVary} we provide the parameters that vary spatially across the study area, along with the distribution that are generated from. 

\begin{longtable}{ | L{1.3cm} L{6cm} L{2cm} L{4cm} |}
\hline 
Symbol & Definition & Value & Source \\ 
\hline \hline
$\lambda$ & Number of bites per mosquito & 10 per month & \cite{Rajagopalan:1980,Subramanian:1994}\\[8pt] 
$a_\text{max}$ & Age at which exposure to mosquitoes reaches its max  level & 20 & \cite{Subramanian:2004} \\[8pt]  
$\psi_1$ & Proportion of L3 leaving mosquito per bite & 0.414 & \cite{Hairston:1968} \\[8pt]  
$\psi_2$ & Proportion of L3 leaving mosquito that enter host & 0.32& \cite{Ho:1967}\\[8pt]  
$s_2$ & Proportion of L3 entering host that develop into adult worms & 0.00275 & \cite{Norman:2000,Stolk:2008}\\[8pt]  
$\mu$ & Death rate of adult worms & 0.0104 per month & \cite{Evans:1993}\\[8pt]  
$\alpha$ & Production rate of Mf per worm& 0.2 per month & \cite{Hairston:1968}\\[8pt]  
$\gamma$ & Death rate of Mf& 0.1 per month& \cite{Hairston:1968,Ottesen:1995}\\[8pt]  
$g$ & Proportion of mosquitoes which pick up infection when biting an infected host& 0.37& \cite{Subramanian:1998}\\[8pt] 
$\sigma$ & Death rate of mosquitoes&5 per month & \cite{Ho:1967} \\[8pt] 
$h(a)$ & Parameter to adjust rate at which individuals of age $a$ are bitten & Linear from 0 to 10, with max 1 & \cite{Norman:2000} \\[8pt] 
$\chi_1$ & Proportion of Mf killed for an individual MDA round using ALB and DEC & 0.95 & \cite{Ismail:1998,Michael:2004}\\[8pt]  
$\kappa_1$ & Proportion of adult worm permanently sterilised during MDA round using ALB and DEC & 0.55 & \cite{Ismail:1998,Michael:2004} \\[8pt] 
$\rho$ &Systematic adherence of MDA & 0.35& \cite{Stolk:2018}\\ \hline
\end{longtable}
\setcounter{table}{0}
\begin{table}[h!]
\vspace*{-0.8cm}
\caption{ Description of the basic model parameters that are assumed to be fixed across the study area. \label{tab:Parameters}}
\end{table}

\begin{table}[!]
\begin{tabular}{ | L{1.3cm} L{6cm} L{6cm} |}
\hline 
Symbol & Definition & Value  \\ 
\hline \hline
$\eta$ & Size of population in simulation & Values are draw from the proposal distribution given in Figure \ref{fig:Distributions}(a) \\[8pt] 
$V/H$ & Ratio of number of vectors to hosts & Values are draw from the joint prior distribution given in Figure \ref{fig:Distributions}(b) \\[8pt] 
$k$ & Aggregation parameter of individual exposure to mosquitoes & Values are draw from the joint prior distribution given in Figure \ref{fig:Distributions}(b) \\[8pt] 
$\alpha_\text{Imp}$ & Importation rate & Values are draw from a uniform prior distribution $\mathcal{U}(0, 0.0005)$ \\[8pt] 
$p_C$ & Coverage of MDA & 65\% or 80\% depending on the future intervention assumptions \\ \hline
\end{tabular}
\caption{ Description of the basic model parameters that are assumed to be varied across the study area. \label{tab:ParametersVary}} 
\end{table}

\newpage
\subsection{Implementation details}
\label{sec:TRANSFIL:Implementation}

To generate the require range of mf prevalences for the seven African countries that we considered, i.e Ethiopia, Sudan, South Sudan, Eritrea, Kenya, Tanzania and Uganda, we varied four parameters of the model; the population size, the vector to host ratio ($V/H$), the aggregation parameter of individual exposure to mosquitoes ($k$) and the importation rate ($\alpha_\text{Imp}$). More specifically, for the importation rate we used a uniform prior distribution with minimum 0 and maximum 0.0005 (max $\frac{5}{10\, 000}$ infections per month). We investigate different maximum values for the importation rate, and we chose the one which give us the desired results without driving the dynamics of the disease. In addition, the interventions reduce the prevalence over time, and therefore as years pass, we decrease the importation rate after intervention in proportion to the reduction in prevalence seen in pilot simulations. More specifically, we produce 2\,000 simulations with constant importation rate over five years of MDA. Then, for our main set of simulations, we adjust the importation rate according to how the prevalence changed in our pilot runs after the intervention was applied.

\begin{figure}[!]
\centering
\subfigure[Proposal density of the population sizes.]{
\includegraphics[scale=0.95]{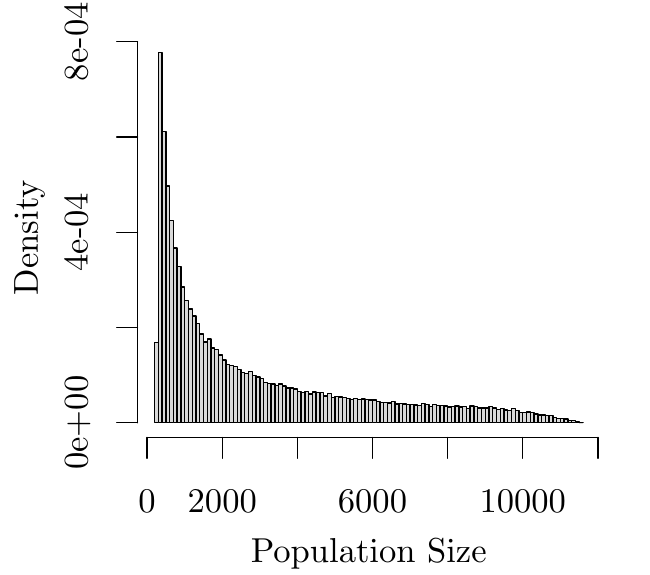} 
\label{fig:ProposalPop}
}
\subfigure[Joint prior distribution on the vector to host ratio ($V/H$) and aggregation parameter of individual exposure to mosquitoes ($k$). \label{fig:Priork_VtoH}]{
\includegraphics[scale=0.95]{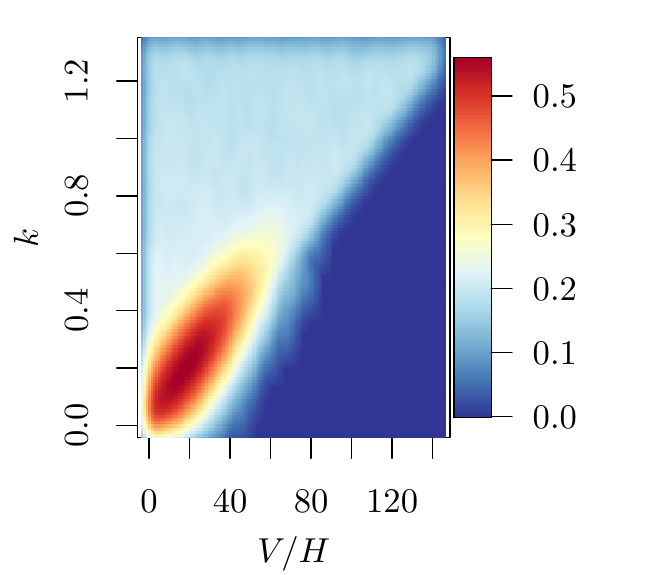} 
}
\caption{Distributions assigned to parameters that vary spatially across the study area. \label{fig:Distributions}}
\end{figure}

Parameters $V/H$ and $k$ were drawn from a range of plausible values based on previously analysed data \citep{Irvine:2015,Irvine:2017,Smith:2017}. The graphical representation of their prior distribution is shown in the left panel of Figure \ref{fig:Distributions}. We then generate 100\,000 parameter vectors by randomly sampled from these prior distributions of parameters $V/H$, $k$ and $\alpha_\text{Imp}$ and from the proposal distribution of the population size as described in Section \ref{subsec:Results} and showed in the right panel of Figure \ref{fig:Distributions}. These samples were then used to generate 100\,000  simulations from the model for each scenario specified in the main text, using the exact years and coverage of the MDA treatments. Our simulations here are focused on areas with anopheles as the dominant vector species. Finally, we set $\delta = 0.01$.

\newpage
\section{Additional results}
\label{sec:Supplementary}

\setcounter{table}{0}
\setcounter{figure}{0}

\begin{figure}[!]
\centering
\subfigure[Comparison of the estimated number of people per pixel with the observed value. \label{fig:Pop}]{
\includegraphics[scale=0.95]{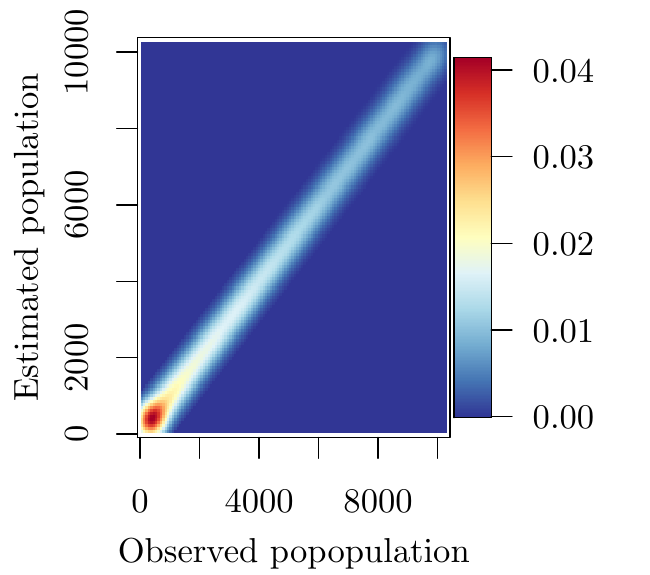} 
}
\subfigure[Effective sample size against the estimated median prevalence for each pixel.]{
\includegraphics[scale=0.95]{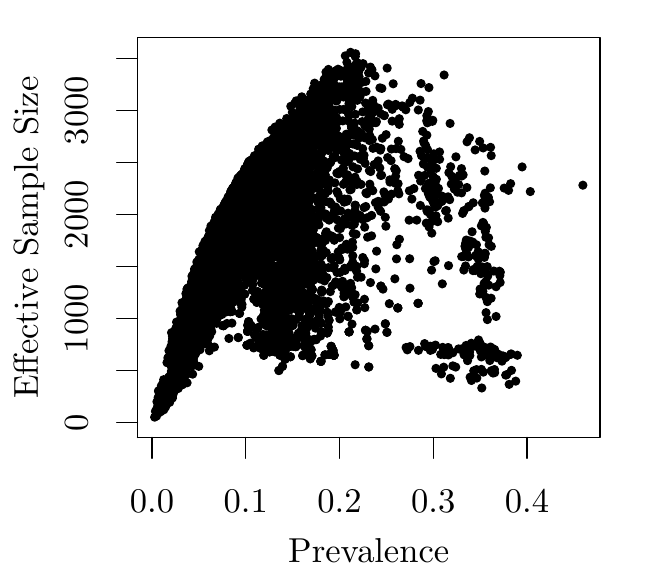} 
\label{fig:ESS}
}
\caption{Performance assessment of our method. \label{fig:PopandESS}}
\end{figure}

\begin{figure}[!]
\centering
	\includegraphics[scale=1]{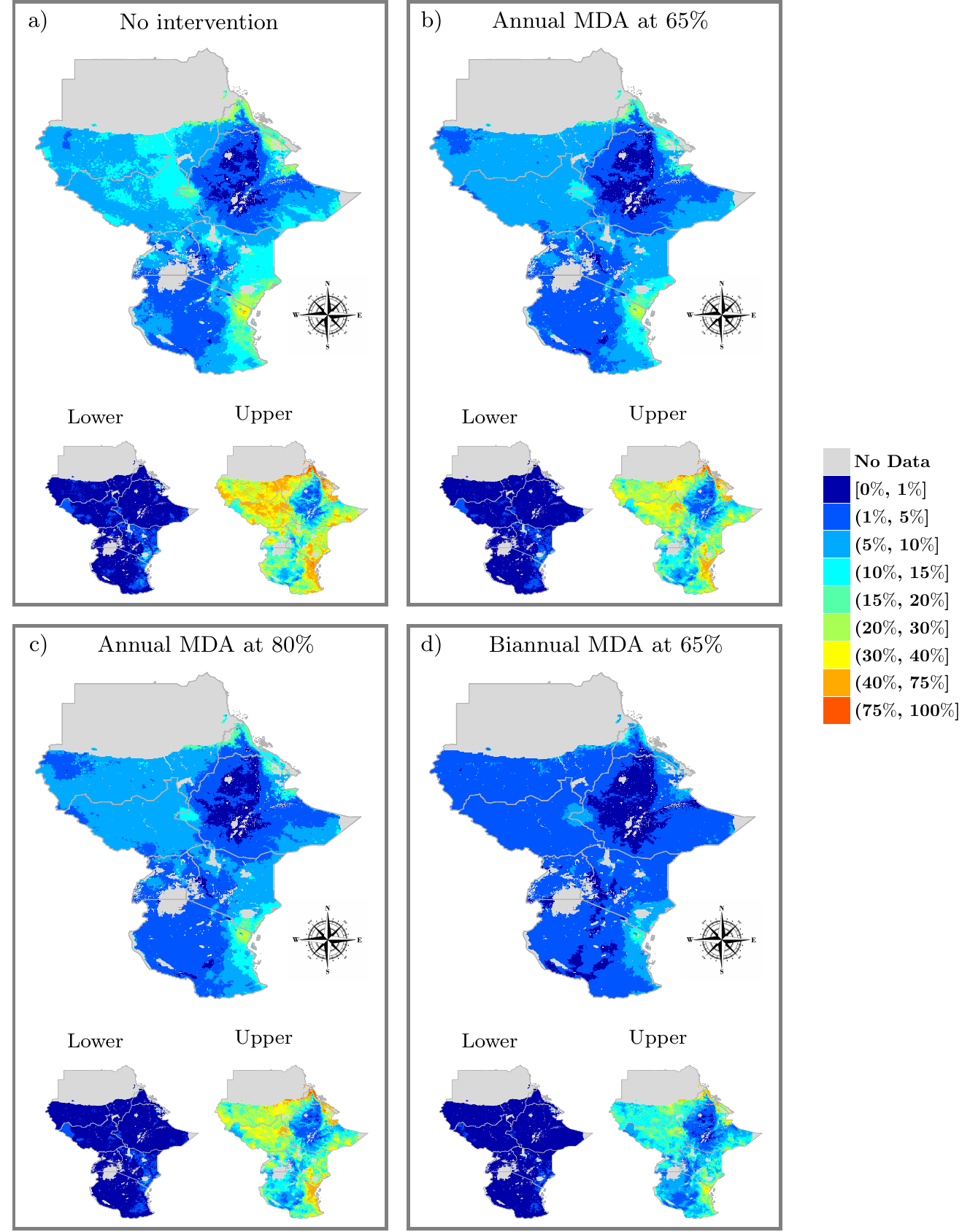}
\caption{Prevalence after 1 year under: a) no intervention; annual MDA with coverage of b) 65\%; c) 80\% and d) biannual MDA at 65\% coverage predicted at $5\times5$ km resolution. Point estimates along with lower (2.5$\%$) and upper (97.5$\%$) percentiles are presented.}
\label{fig:PrevYear1}
\end{figure}

\begin{figure}[!]
\centering
	\includegraphics[scale=1]{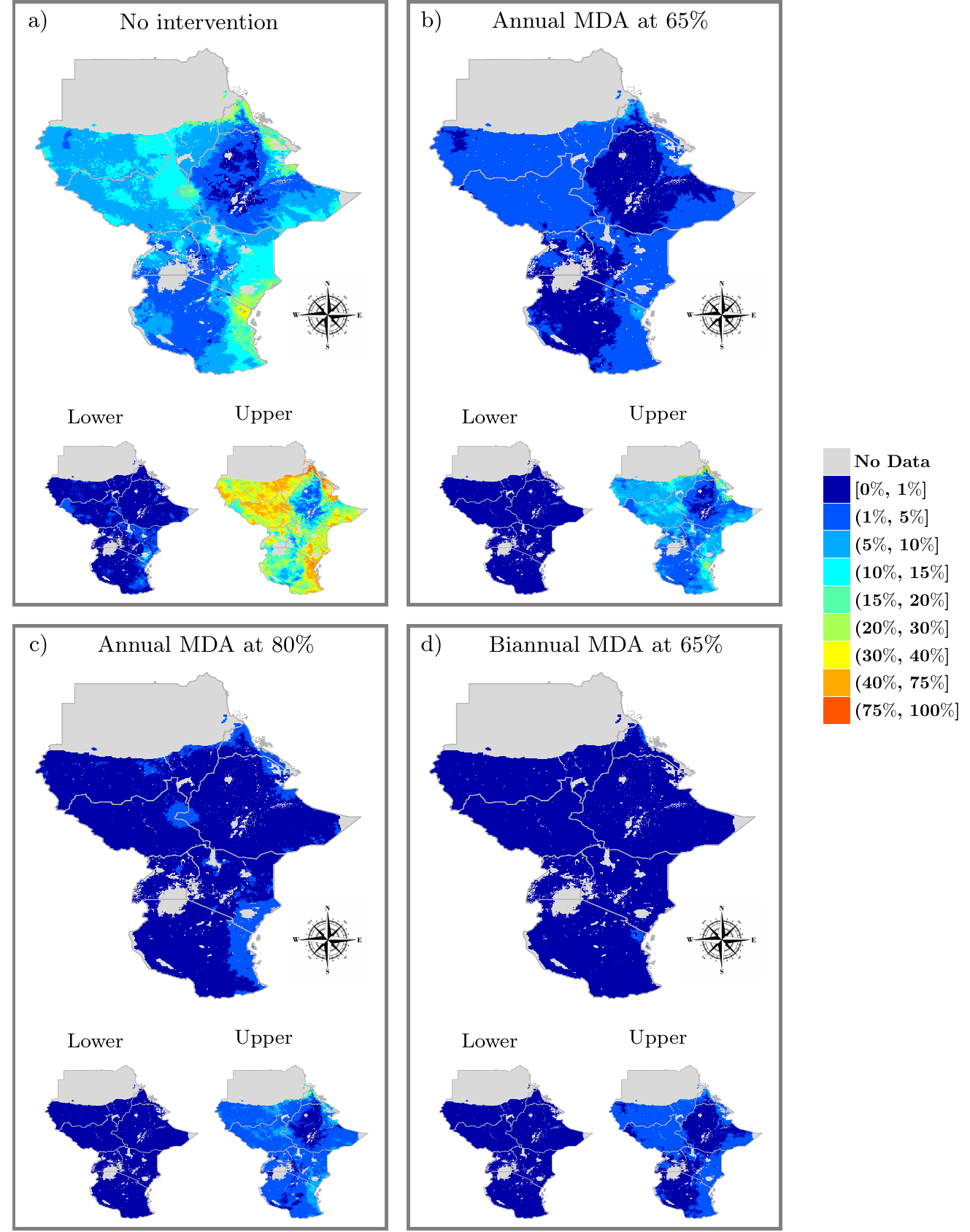}
\caption{Prevalence after 5 years under: a) no intervention; annual MDA with coverage of b) 65\%; c) 80\% and d) biannual MDA at 65\% coverage predicted at $5\times5$ km resolution. Point estimates along with lower (2.5$\%$) and upper (97.5$\%$) percentiles are presented.}
\label{fig:PrevYear5}
\end{figure}

In this section we provide additional results for the analysis of the LF data in East Africa of Section \ref{subsec:Results} of the main manuscript. The performance of the method described in Section \ref{sec:Methods} was assessed by comparing the observed and the estimated number of people at each pixel in the left panel of Figure \ref{fig:PopandESS}. In the right panel of Figure \ref{fig:PopandESS}, we further evaluated the performance of our method by plotting effective sample size per pixel. Figures \ref{fig:PrevYear1} and \ref{fig:PrevYear5} illustrate future predictions of the prevalence under different control scenarios for the first and fifth year.

\clearpage
 \bibliographystyle{agsm}

\end{document}